\documentclass{aa}

\bibpunct{(}{)}{;}{a}{}{,} 
\usepackage{graphicx}

\usepackage{txfonts}
\usepackage{multirow}
\usepackage{booktabs}

\begin{document}

   \title{A deeper solution to the actual geometry of CCD mosaic chips}

   \subtitle{}

   \author{Z.J.~Zheng
          \inst{1,4},
          Q.Y.~Peng\inst{1,4}\thanks{Contact:tpengqy@jnu.edu.cn}
          \and
          A. Vienne\inst{2,4}
          \and
          F.R.~Lin\inst{3,4}
          \and
          B.F.~Guo\inst{1,4}
          }

   \institute{Department of Computer Science, Jinan University, Guangzhou 510632, P. R. China\\
         \and
          IMCCE, Observatoire de Paris, UMR 8028 du CNRS, UPMC, Universit\'{e} de Lille, 77 av. Denfert-Rochereau, 75014 Paris, France\\
         \and
         School of Software, Jiangxi Normal University, Nanchang 330022, P. R. China\\
         \and
         Sino-French Joint Laboratory for Astrometry, Dynamics and Space Science, Jinan University, Guangzhou 510632, P. R. China
         }

   \date{}


  \abstract
   {For charge-coupled device (CCD) mosaic chips in the focal plane of a large telescope, the unification for all the measurements of each chip is vital to some scientific projects, such as deep astrometric standards or construction for deeper images that can also seamlessly cover a larger area of the sky. A key part of the reduction involves the accurate geometric distortion~(GD) correction and the precise determination of the relative positions of the CCD chips. The short-term and long-term stabilities of them are also important when it comes to studying whether there are systematic variations in the optical system of the telescope.}
   {We present a solution to determine the actual or physical relative positions between CCD chips. Due to the limited depth of the Gaia catalogue, there may be few stars identified from the Gaia catalogue for astrometric calibration on the deep observation of a large, ground-based or space-based telescope, such as the planned two-metre Chinese Space Station Telescope~(CSST). For this reason, we referred to the idea from the Hubble Space Telescope (HST) astrometry to only use stars' pixel positions to derive the relative positions between chips. We refer to the practice as differential astrometry in this paper. In order to ensure the results are reliable, we took advantage of Gaia EDR3 to derive the relative positions between chips, to provide a close comparison. We refer to the practice as photographic astrometry.}
   {By taking advantage of the GD solution and the common distortion-free frame derived from the observations, we related the physical positions of the adjacent pixel edges of two CCD chips and estimated the actual relative positions between chips. We implemented the technique for the CCD mosaic chips of the Bok 2.3-m telescope at Kitt Peak based on two epochs of observations~(Jan 17,2016 and Mar 5,2017).}
   {There is a good agreement between the two types of astrometry for the relative positions between chips. For the two epochs of observations, the averages of the gaps derived from photographic astrometry and differential astrometry differ to about 0.046 pixels~($\sim$0.021 arcsec) and 0.001 pixels~(<0.001 arcsec), respectively, while the average precisions of the gaps are about 0.018 pixel~($\sim$0.008 arcsec) and 0.028 pixels~($\sim$0.013 arcsec), respectively. The results provide us with more confidence in applying this solution for the CCD mosaic chips of the CSST by means of differential astrometry. Compared with the solution described by Anderson \& King, which has been used to determine the interchip offset of Wide Field Planetary Camera 2~(WFPC2) chips and Wide Field Camera 3~(WFC3) chips at the HST, the solution proposed in this paper shows at least a factor of two improvement in precision, on average.}
   {We think there are two definite advantages of our method. On one hand, we perform the measurements for two adjacent edges instead of two individual chips, allowing the results to be as local as possible, and meanwhile we alleviate the propagated error of residual distortions of each observation deviating from the average solution throughout the field of view~(FOV). On the other hand, the final outcome is not mixed up with GD effects, which would bias the realistic geometry of the CCD mosaic chips. Therefore, the proposed method is expected to be an effective technique to monitor the stability of the CCD mosaic chips in the CSST and other ground-based CCD mosaic as well.}

\titlerunning{A solution to determine the geometry of CCD mosaic chip}
   \maketitle

\section{Introduction}

Modern telescopes are built with larger and larger apertures to enhance detection capability for searching dim targets. However, to fully exploit the scientific productivity of the telescope, an attempt to build a larger single CCD chip would face tough challenges in manufacturing. An alternative technique, called mosaic, is to build a close-packed array, in which a certain number of small-scale CCD chips are placed as close together as possible on a common base or plate, which enables the ground-based and space-based telescope to image a much larger sky area. Moreover, there is a definite trend to design complex mosaic instruments with increasing numbers of chip elements~\citep{Sekiguchi1992,Gunn1998,DePoy2008,Lou2016}.
Notwithstanding, a larger mosaic size brings additional difficulties in data processing. Since each CCD chip in the mounting plate is an autonomous detector and their properties vary in quantum efficiency, readout noise, dark current, and so on, careful calibration is required before the unification of the measurement of each CCD chip~\citep{Luppino1995}. For high-precision astrometry, two key factors that limit the accuracy and precision of the unification are (1)~geometric distortion~(GD) calibration and (2)~determination of relative positions of the CCD chips. For example, with the application of the GD solution for WFPC2's corrective optics, as well as the changes in the relative positions of the four WFPC2 chips~(including three WF chips and one PC chip), over time, Saturn satellite astrometry from multiple chips is combined to achieve more precise result~\citep{French2006} than when using only PC images. Some projects such as the combination of different types of images especially obtained with different chips for the construction of deeper images also rely on the high accuracy of these solutions~\citep{Wadadekar2006}.

A regular check for the stability of CCD mosaic geometry is necessary, especially before and after a noticeable change in the thermal environment of an optical system, since the derived calibration parameters in lab tests are not proven to be appropriate or stable for on-sky measurement. For example, a significant variation for the metrics of the NOAO CCD Mosaic Imager indicates possible thermal cycling of a dewar, which results from the difference in thermal expansion coefficients between silicon and copper~\citep{Platais2002}. Another example is that the relative motion of CCDs over just one week can even be as large as 200 mas, nearly a pixel for the Dark Energy Camera, as a result of a warming or cooling event for the camera~(see Fig. 15 in~\citealt{Bernstein2017}). As one of the best space telescopes, the Hubble Space Telescope~(HST) is designed for such precise positional imaging astrometry that a small variation of its optical system even over one orbital period should be taken into account or monitored. As the HST orbits the Earth while the Earth orbits the Sun, its focus variations are correlated with thermal variations induced by the angles between the Sun and the telescope~(the so-called breathing effect). Also, a clear monotonic trend for the interchip offset of the WFPC2 chips was found by~\cite{Anderson2003}.

The two-metre Chinese Space Station Optical Survey is a planned full sky survey operated by the Chinese Space Station Telescope~(CSST). The CSST will be equipped with five first-generation instruments including a survey camera, a terahertz receiver, a multi-channel
imager, an integral field spectrograph, and a cool planet imaging coronagraph~\citep{ZhanHu2021}. The survey camera is the most important instrument; its aim is wide-area multi-band imaging and slitless spectroscopic survey, and its field of view (FOV) is 1.1${\times}$1.2 deg$^{2}$. It is equipped with 30 9k${\times}$9k CCD detectors, of which 18 CCDs are for imaging observation and 12 CCDs are for spectroscopic observation.

Before the CSST begins science operations around 2024, we should have technical preparation for monitoring the interchip offset of its CCD mosaic chips, for testing the optical system's stability. In this paper, the CCD mosaic camera of the Bok 2.3-m telescope, which includes 4 4k$\times$4k CCDs, is used as a preliminary prototype of the CSST's CCD mosaic chips to test the solution to the geometry of CCD mosaic chips proposed by \cite{Wang2019}.

Although \cite{Wang2019} derived the relative positions between chips of the Bok 2.3-m telescope by using Gaia DR2, we refer to the idea from the astrometry of the HST and propose an alternate solution for the relative position between chips by only using stars' pixel positions. It is more practical for the ground-based or space-based deep observations such as the CSST, which can observe much fainter stars~(e.g. 26 mag in g band) than the Gaia satellite, and we can usually only obtain their pixel positions. In this situation, there are few stars identified from the Gaia catalogue for astrometric calibration. In order to ensure the results were reliable, we took advantage of Gaia EDR3 to derive the relative positions between chips as in \cite{Wang2019} in order to provide a close comparison. For simplicity, we refer to the practice of using Gaia EDR3 to derive the relative positions between chips as photographic astrometry in this paper. The practice of only using stars' pixel positions is referred to as differential astrometry.

During the reduction, we tried our best to eliminate the possible errors on the basis of \cite{Wang2019}. Firstly, we corrected the offset in altitude introduced by the differential colour refraction effects. In addition, we adopted some criteria to reject sources with possible bad astrometry based on Gaia EDR3's re-normalised unit weight error~(RUWE), and we also rejected sources that suffered from saturated or crowding status on the observations. We also introduced the weighted scheme to reduce the effect of a large number of low signal-to-noise (S/N) stars on the reduction.

This paper is organised as follows. In Sect.~2, we briefly describe the instrument and the observations. In Sect.~3, we introduce the procedure of deriving the GD solution when using photographic or differential astrometry. In Sect.~4, we elaborate the solution for relative positions between the CCD chips, based on photographic astrometry and differential astrometry, respectively. The results are shown and discussions are made in Sect.~5. In Sect.~6, we compare the solution for the relative positions between chips proposed in this paper with the solution described by Anderson \& King~(\citeyear{Anderson2003}, hereafter the AK03 method), which has been used to determine the interchip offset of WFPC2 chips and WFC3 chips~\citep{Bellini2009} at the HST. The conclusion is drawn in the last section.

\section{Instrument and observations}

The observations are only taken to derive the GD solution for the 90Prime camera of the Bok 2.3-m telescope when it serves the Beijing-Arizona Sky Survey~(BASS). Although more detailed information about the BASS project can be found in \cite{Wang2019}, we briefly summarise the project below.

\begin{figure}[htp]
\centering
\includegraphics[width=\columnwidth, angle=0]{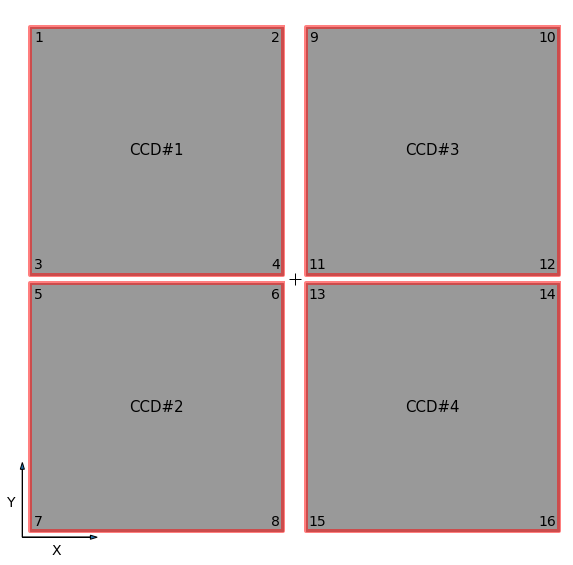}
\caption{CCD mosaic layout in the 90 Prime camera of the Bok 2.3-m telescope. There are four CCDs:
CCD\#1, CCD\#2, CCD\#3, and CCD\#4, and four identifiers are listed at the corners of each CCD. For CCD\#1, the pixel position of Corners 1, 2, 3, and 4 are (1,4032), (4096,4032), (1,1), and (4096,1), respectively. The pixel positions of corners in the other chips are the same as CCD\#1.}
\label{Fig2}
\end{figure}

The BASS is a new imaging survey resulting from the collaboration between the National Astronomical Observatory of China (NAOC) and Steward Observatory. It uses the Bok 2.3-m telescope, which is located at Kitt Peak~(IAU code 691, E248${\degr}$ 24${\arcmin}$ 3.6${\arcsec}$, N31${\degr}$ 57${\arcmin}$ 28.7${\arcsec}$) with the 90Prime camera installed at the prime focus~(4 CCDs, 4096 $\times$ 4032 pixels each, as shown in Fig.~\ref{Fig2}) to survey an area of about 5400 deg$^{2}$ in the north Galactic cap. The photometric system includes the SDSS g and DES r bands. The expected depths for 5$\sigma$ point sources and Galactic extinction correction are 24.0 mag for the g band and 23.4 mag for the r band, respectively.

In order to push the astrometric capability of the Bok 2.3-m telescope to its limits, the GD solution is needed. We attain two epochs of observations to derive the GD solution. Specifically, a dithered observational scheme is organised in an array of 7${\times}$7 pointings with an offset of about 10${\arcmin}$~\citep{Wang2019}. The exposure time of each observation is 60.0 s. The specifications of the instrument and the observations are given in Table~\ref{Tab1} and Table~\ref{Tab2}. For simplicity, the two observation-sets are called Obs16~(January 17, 2016) and Obs17~(March 5, 2017), respectively.

It should be noted that, due to lack of experience, Obs16 was centred on a sparse field~(coordinates $(\alpha,\delta)_{\rm  J2000.0}\sim(\rm
10^h48^m12^s\!\!.0,+28^{\circ}45^{\prime}46^{\prime\prime}\!\!.9)$). There are only about 300 stars on each chip, and they are inadequate to derive a fine GD solution. Also, the observation was interrupted on that night. About one hour later, the observation resumed. Considering the distortion is continuously evolving, only the last 30 frames~(sorted by observation time) of Obs16 were chosen for processing in this paper. In 2017, a much denser calibration field was centred on coordinates $(\alpha,\delta)_{\rm  J2000.0}\sim(\rm
06^h48^m12^s\!\!.0,+00^{\circ}45^{\prime}47^{\prime\prime}\!\!.0)$. There are about 3000 stars on each chip. Because of the cloudy weather, we discarded 11 frames of Obs17 with poor image quality.

\begin{table}
\caption{Specifications of 90Prime camera in the Bok 2.3-m telescope.} \centering
\begin{tabular}{rr}
\hline\hline\noalign{\smallskip}
F-Ratio                           &f/2.98\\
CCD field of view            &1.08${\degr}$${\times}$1.03${\degr}$\\
Absolute pointing                &$<$ 3${\arcsec}$ RMS \\
Size of CCD array            &4096${\times}$4032${\times}$4\\
Size of pixel   & $15{\times}15$ $\mu$m${^2}$\\
Approximate pixel scale                 & $0\farcs453$/pix \\
\hline
\end{tabular}
\label{Tab1}
\end{table}

\begin{table}
\caption{Specifications of the observations.}
\centering
\begin{tabular}{@{}ccccccc@{}}
\hline\hline\noalign{\smallskip}
Date  & Abbr. & No.  & Filter  & Seeing &  Airmass  \\
\hline\noalign{\smallskip}
Jan 17,2016 & Obs16 & 32 & DES r & 1.7${\arcsec}$-2.4${\arcsec}$ & 1.05-1.14 \\
Mar 5,2017 & Obs17 & 38 & SDSS g & 1.3${\arcsec}$-1.7${\arcsec}$ & 1.16-1.20 \\
\hline
\end{tabular}
\tablefoot{The second column lists the abbreviation for the observation set. The third column lists the number of frames in each observation set. The fourth column lists the adopted filter. The last two columns list the seeing and the range of the air mass for the observations, respectively.}
\label{Tab2}
\end{table}

\section{The GD solution}

This section mainly describes the GD solutions for the observations by means of two types of astrometry. For simplicity, the GD solution through photographic astrometry is called GD1, while the GD solution via differential astrometry is called GD2.
\subsection{Data pre-processing}

To derive an accurate GD solution, we first performed a two-dimensional Gaussian centring to obtain the pixel positions of the star images on the frame. After that, we cross-matched the pixel positions with the sources in the Gaia EDR3, considering their motions and gnomonic projection effects.

It should be noted that fewer sources are found in Gaia EDR3 than Gaia DR2~\citep{Gaia2018} for the field of Obs17, which is in a very crowded area. According to~\cite{Riello2021}, in the data processing of Gaia EDR3, the background treatment for BP and RP has been considerably improved to avoid systematic problems caused by crowding, and as a result, the crowding evaluation in Gaia EDR3's processing leads to a cleaner list of internal calibration sources.
We then removed many sources with possible bad astrometry, whose RUWEs in Gaia EDR3 are larger than 1.4 for Obs16 and 1.1 for Obs17, respectively, determined by the numbers of the observed star in each frame and the suggestion given by the Gaia Collaboration~(\citeyear{Gaia2021}). We also excluded stars that are saturated or suffer from the contamination of blended neighbour(s) or saturated star(s) on the observation. Finally, we obtain about 3,000 Gaia stars for Obs16 and about 120,000 for Obs17. The above processings are based on the software developed by us~\citep{Peng2012,Peng2015MNRAS,Peng2017,Wang2017}.

\subsection{Differential colour refraction correction}

To achieve high-accuracy results through photographic astrometry, we considered the differential colour refraction~(DCR) especially in the observation of Obs17, which is observed at $\approx$30${\degr}$ zenith distance with the SDSS g filter. However, the observations were not taken with several filters on one night, so we cannot correct the DCR using the colour index of stars by the photometry of them. We referred to the technique proposed by \cite{Lin2020}, which corrects the DCR of each filter against the colour index proposed by catalogue. Since the filters of the Pan-STARRS1~(PS1) survey are similar to the filters used for the observations, we cross-identified sources between PS1 and Gaia EDR3 to obtain the stars' $gmag$~(central wavelength 4866${\AA}$) and $rmag$~(central wavelength 6215${\AA}$) provided by PS1. They correspond to the SDSS g filter~(central wavelength 4776${\AA}$) and the DES r filter~(central wavelength 6412${\AA}$) used by the BASS, respectively.

After we cross-matched the observations with Gaia EDR3 and PS1 data, we performed the reduction in the alt-azimuth coordinate system with a fourth-order polynomial, considering all astrometric effects including proper motion, parallax, aberration, atmosphere refraction, and projection effects. It should be noted that we did not take the \textit{RA-OBS} and \textit{DE-OBS} keywords recorded in the FITS header as the tangential point as \cite{Wang2019} did, since they are J2000.0 positions and would affect the accurate calculation for atmospheric refraction. Instead, we calculated their apparent R.A. and Dec. positions and transformed them to the apparent azimuth and altitude positions. For atmosphere refraction correction, a simple $A$\,tan\,$z$+$B$\,tan\,$^{3}z$ model ($z$ is the zenith distance) provided by SOFA library\footnote{\url{http://www.iausofa.org}} was adopted.

\begin{figure*}[htp]
\centering
\includegraphics[width=\textwidth, angle=0]{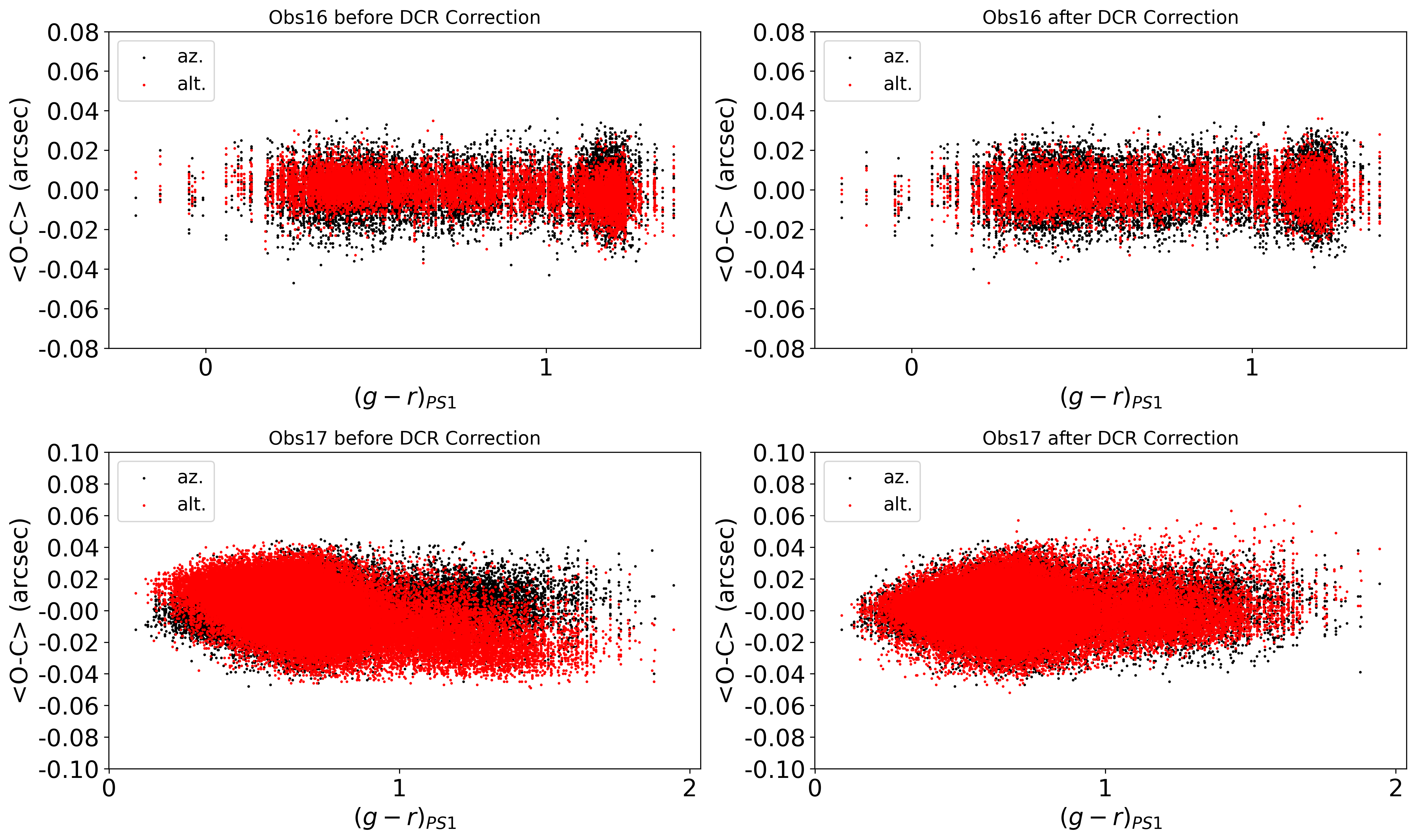}
\caption{Residual $(O-C)$ in alt-azimuth coordinate system against $(g-r)_{PS1}$ before and after DCR correction. The top two panels are $\langle O-C\rangle$ before and after DCR correction for Obs16. The bottom two panels are $(O-C)$ before and after DCR correction for Obs17. We note that the results of Obs17 are based on the stars brighter than 18 Gmag and the results of Obs16 are based on the stars brighter than 17 Gmag.}
\label{FigDCR}
\end{figure*}

The fourth-order polynomial makes it possible to absorb the GD effects in the observations. The residuals in altitude mainly reveal the DCR of stars. We adopted Eq. 5 of \cite{Lin2020} to derive the following DCR solution:
$$
\operatorname{DCR}\left(a_{1}, a_{2}\right)=a_{1}+a_{2}\cdot color\cdot \tan z,
$$
where $color$ and $z$ are the colour index ($gmag-rmag$) and the zenith distance of the star, respectively, and $a_{1}$ and $a_{2}$ are the fitted DCR parameters. The DCR solutions of each chip are shown in Table~\ref{TabDCR}. After DCR correction, there are no significant systematic errors in the residual~($\langle O-C\rangle$) in altitude against $(g-r)_{PS1}$, especially for Obs17, as shown in Fig.~\ref{FigDCR}. We note that the results of Obs17 are based on the stars brighter than 18 Gmag
and the results of Obs16 are based on the stars brighter than 17 Gmag. The dispersion of the $\langle O-C\rangle$s in altitude is nearly the same as that in the unaffected azimuth direction.

\begin{table}
\caption{DCR solution for the observations.}
\centering
\begin{tabular}{@{}cccc@{}}
\hline\hline\noalign{\smallskip}
Obs-set  & CCD\# & $a_{1}$  & $a_{2}$  \\
\hline\noalign{\smallskip}
\multirow{4}*{Obs16} & 1 & 0.006 & -0.013  \\
 & 2 & 0.005 & -0.016  \\
  &3 & 0.008 & -0.022 \\
   & 4 & 0.007 & -0.020 \\
\hline\noalign{\smallskip}
  \multirow{4}*{Obs17} & 1 & 0.019 & -0.045  \\
 & 2 & 0.028 & -0.060 \\
  & 3 & 0.019 & -0.046 \\
   &4 & 0.028 & -0.059 \\
\hline
\end{tabular}
\tablefoot{The first column lists the observation set. The second column lists the index of the CCD chip. The last two columns list the coefficients of the solution. The units are in arcsec.}
\label{TabDCR}
\end{table}

\subsection{GD1 procedure by means of photographic astrometry}

The most direct way to solve the GD solution is to compare the observed positions with the positions of a distortion-free frame~(DFF). By definition~\citep{Anderson2003}, the DFF is a set of positions, into which any observed positions after GD correction can be transformed. The set has nothing more than a displacement, a rotation, and a scale factor from a four-parameter conformal transformation.

Now, Gaia EDR3 provides us with accurate enough positions to build a DFF for the observations. For a faint source (Gmag~20), the expected median error propagated to the astrometric position in J2000.0 of Obs17 is about 0.9 mas, while the uncertainties of the positional centring for a high S/N star image is usually 1$\%$ pixel, which is about 5 mas. So, the positions calculated from Gaia EDR3 can serve as a distortion-free frame for the observations in this paper.
However, we have to consider the possible systematic errors caused by spurious astrometric solutions in Gaia processing. They are mainly due to close source pairs, especially in crowded fields, which even Gaia can not completely resolve~\citep{Fabricius2021}. Although Gaia EDR3 has better angular resolution than Gaia DR2, 1.6$\%$ of the astrometric solutions may be still spurious, which produces meaningless proper motions and parallaxes~\citep{Fabricius2021}. Thanks to the dithered scheme, which allows the same stars to be imaged in different positions of the CCD detectors, their possible systematic errors could be cancelled by implementing the GD1 procedure as described in the following.

For each observation, we took the standard coordinate calculated from Gaia EDR3 as its DFF, considering all astrometric effects (proper motion, parallax, aberration, atmosphere refraction, and differential colour refraction) and projection effects. After gnomonic projection, a four-parameter conformal transformation was used to relate stars' standard coordinates to their pixel positions. The residual between the observed and the calculated positions~($O$ -- $C$) can be resolved into three sources: the GD effect as a function of the position of the stars in the CCD detector, the possible catalogue systematic error as described above, and the measurement error. As the same star in dithered observations falls in different pixel positions on the CCD chip, the GD effect in specific pixel positions can be derived by cancelling out the catalogue's systematic error and compressing the measured errors~(see~\citeauthor{Peng2012} 2012 for details).

Since the field of Obs17 is much denser than Obs16, we can derive a finer GD solution for Obs17. We divided the FOV of each detector into 16$\times$16 elements for Obs16 and 32$\times$32 elements for Obs17, according to the compromise between the need for an adequate number of grid elements to model the GD solution and an adequate sampling of stars~($\geq30$, determined empirically) in each grid element. The GD effect in the centre of each element was estimated from the
average GD effects of all the stars in this element through a $\sigma$-clipped average solution. Solving for the GD solution is an iterative process, with continuously improving positions and reduced residuals. Convergence was reached until the average residual in each element was not larger than 0.001 pixels.

After the GD solution was adopted for the observations, we calculated the standard deviation~($\sigma$) of its residuals for each star. Again we derived a new GD solution using a weighted four-parameter conformal transformation. The weight of each star was set to $1/\sigma^{2}$.

We note that, during the procedure, a third-order polynomial was used to present the large-scale GD, and a look-up table was used to present the small-scale GD. The third-order polynomial is as follows:
$$
\left\{
\begin{array}{rcl}
 \Delta x&\!\!\!=\!\!\!&  \sum\limits_{0\leqslant i+j \leqslant 3} a_{k}\tilde{x}^{i}\tilde{y}^{j}, k=1 \cdots 10   \\
 \Delta y&\!\!\!=\!\!\!&  \sum\limits_{0\leqslant i+j \leqslant 3} b_{k}\tilde{x}^{i}\tilde{y}^{j},
 k=1 \cdots 10,\\
\end{array}
\right.
$$ where $\tilde{x}_{i,j,k}$ and $\tilde{y}_{i,j,k}$ are the
normalised positions, defined as
$$
\left\{
\begin{array}{rcl}
\tilde{x}=(x-2048)/2048 \\
\tilde{y}=(y-2016)/2016. \\
\end{array}
\right.
$$ The normalised positions make it easier to recognise the magnitude
of the contribution of each term at the edge of the chip.
The coefficients of the polynomial are given in Appendix A. The details of the look-up table are not given in this paper, but the interested reader may wish to contact the corresponding author for more information.

The final derived GD1 solutions are shown in the left panel of Fig.~\ref{FigGD}. The median~(MED) of GD1 solutions is about 0.60 pixels smaller than that of~\cite{Wang2019} for Obs16 and 0.68 pixels smaller for Obs17. The maximun~(MAX) GD1 solution is about 1.98 pixels larger than~that of \cite{Wang2019} for Obs16 and 3.03 pixels larger for Obs17. We think the main cause is that we exclude a large amount of sources, which may introduce systematic errors. The introduced weighted scheme reduces the effect of a large number of low S/N stars in the GD1 solution.

\subsection{GD2 procedure by means of differential astrometry}
For differential astrometry, its DFF (master frame in this paper) is constructed directly from the pixel positions of the stars, which the GD-corrected observations can be transform into, with nothing more than a four-parameter conformal transformation. However, as the FOV of the instrument is rather large, the projection effects should be taken into account. Therefore, we constructed a local master frame for each observation rather than constructing a global master frame for all of the observations~\citep{Anderson2003}. The local master frame represents a set of distortion-free positions, which can be observed with the used chip at corresponding pointing.

Specifically, when a single observation was taken as the root of the local master frame, only other observations that have enough common stars~($\geq$30, determined empirically) with the root observation, were chosen for the construction of the local master frame. The constraint reduces the projection effects on the positions of the master frame, since they were derived from the observations whose pointings are not too far away from the pointing of the root observation.

In detail, we found a conformal transformation from each observation to the root observation by the least-squares method, determined from their common stars' pixel positions. The position of each star on the master frame was calculated based on the $\sigma$-clipped average of all the transform positions from the same stars of different observations. The standard deviation of each position of the master frame served as a weight for transformation in the next iteration. At the same time, a star on the frame would be flagged if it had a transform outlier during the estimation for the master frame's position, so the outlier would not be used in the next iteration. Also, new stars were added to the master frame, and, finally, the master frame covers all the observations if they have common stars with the root observation. The primitive master frame was then used to improve the transformation, with the weights and flags determined by the previous iteration. The procedure was iterated several times until the positions on the master frame converged to 0.001 pixels~($\sim$0.5 mas) in each direction. Only the transformation residuals of the common stars were taken as the GD effects and used to derive the GD solution as previous. And after the primitive GD solution was derived, the master frames were reconstructed based on the corrected positions of the observations. The GD2 procedure iterated until the average residual in each element was not larger than 0.001 pixels.

\begin{figure*}[htp]
    \begin{minipage}{\textwidth}
    \centering
        \includegraphics[width=0.48\textwidth, angle=0]{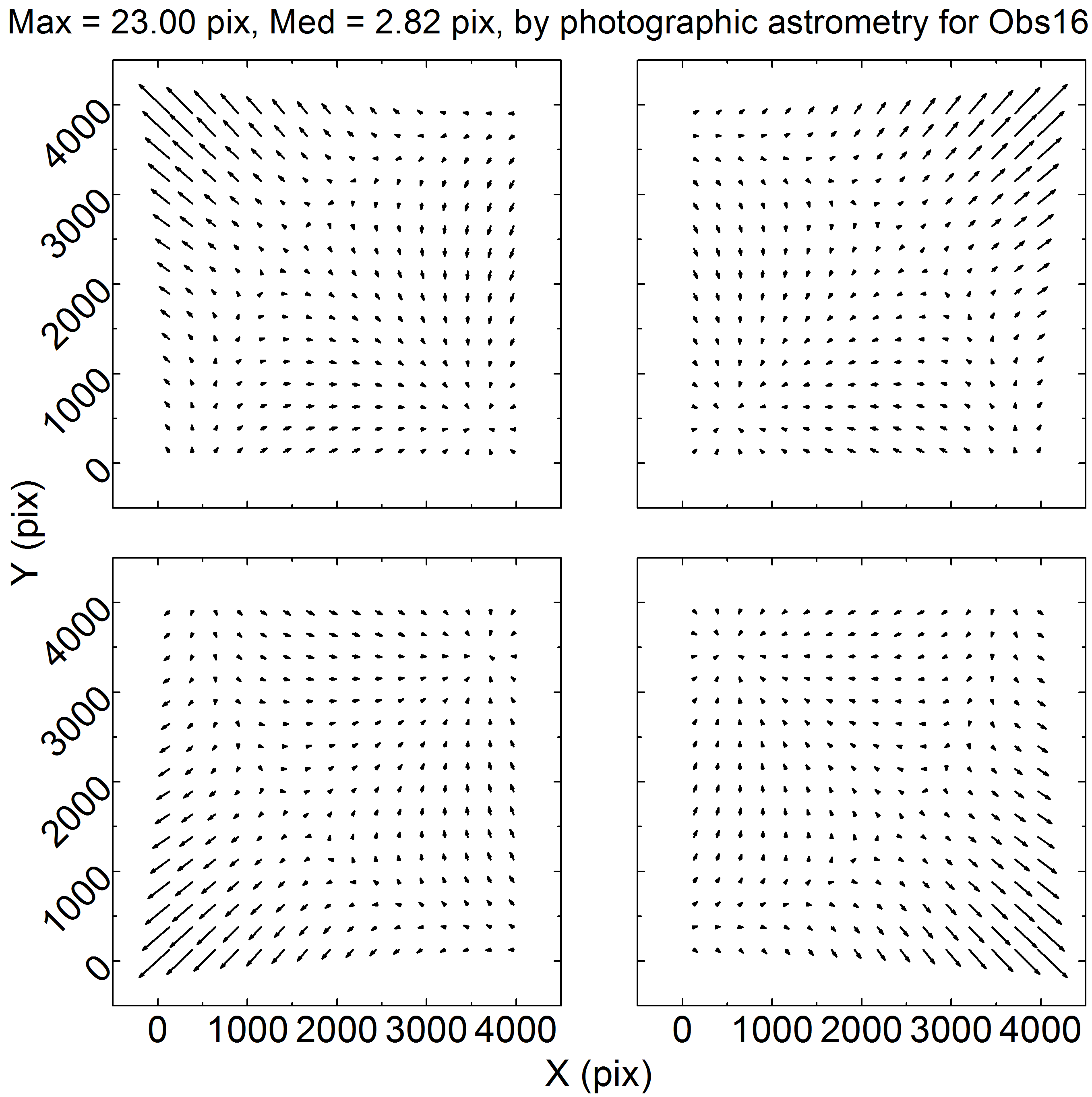}
        \includegraphics[width=0.48\textwidth, angle=0]{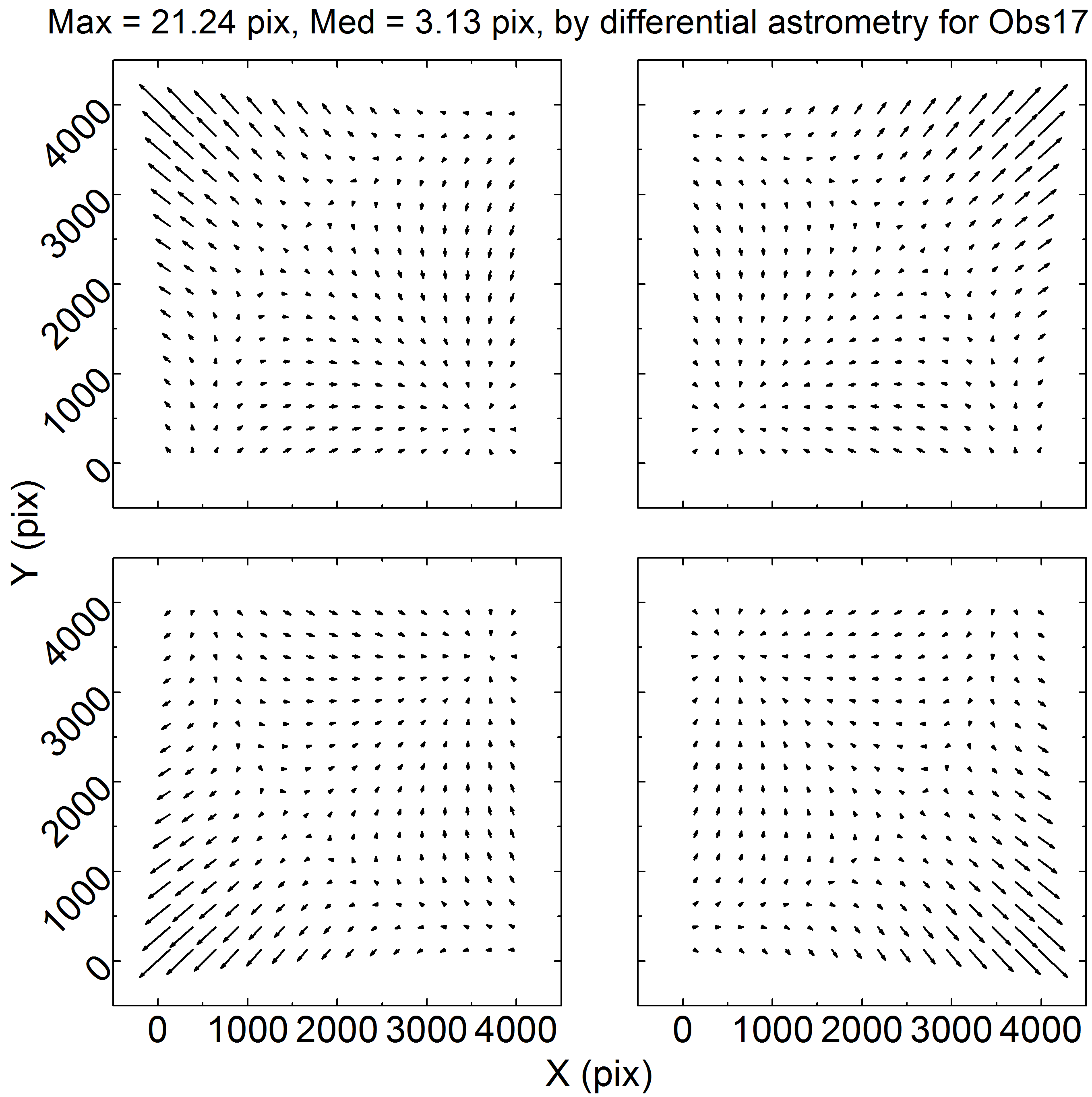}
    \end{minipage}
    \vspace{0in}
    \begin{minipage}{\textwidth}
    \centering
        \includegraphics[width=0.48\textwidth, angle=0]{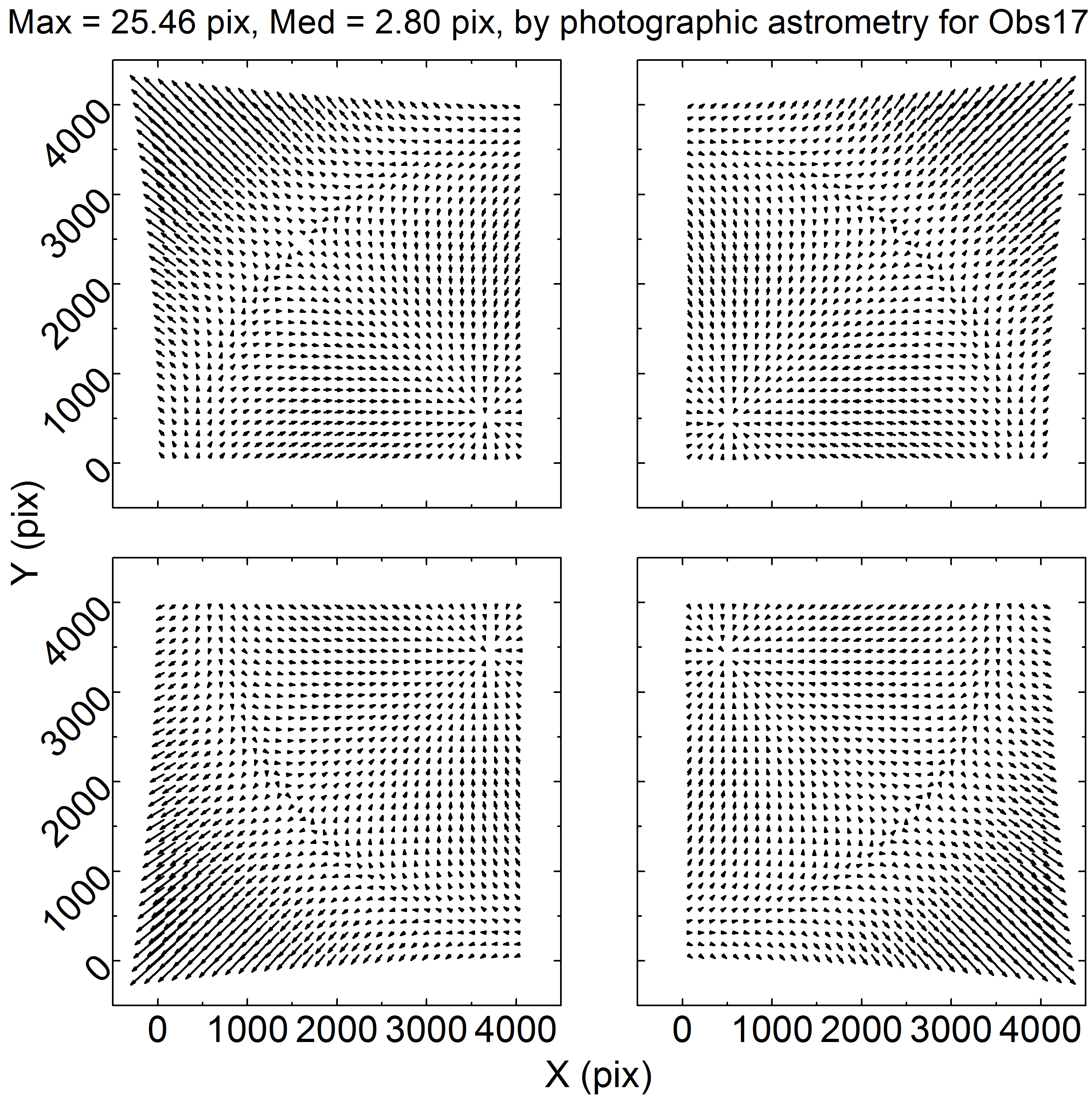}
        \includegraphics[width=0.48\textwidth, angle=0]{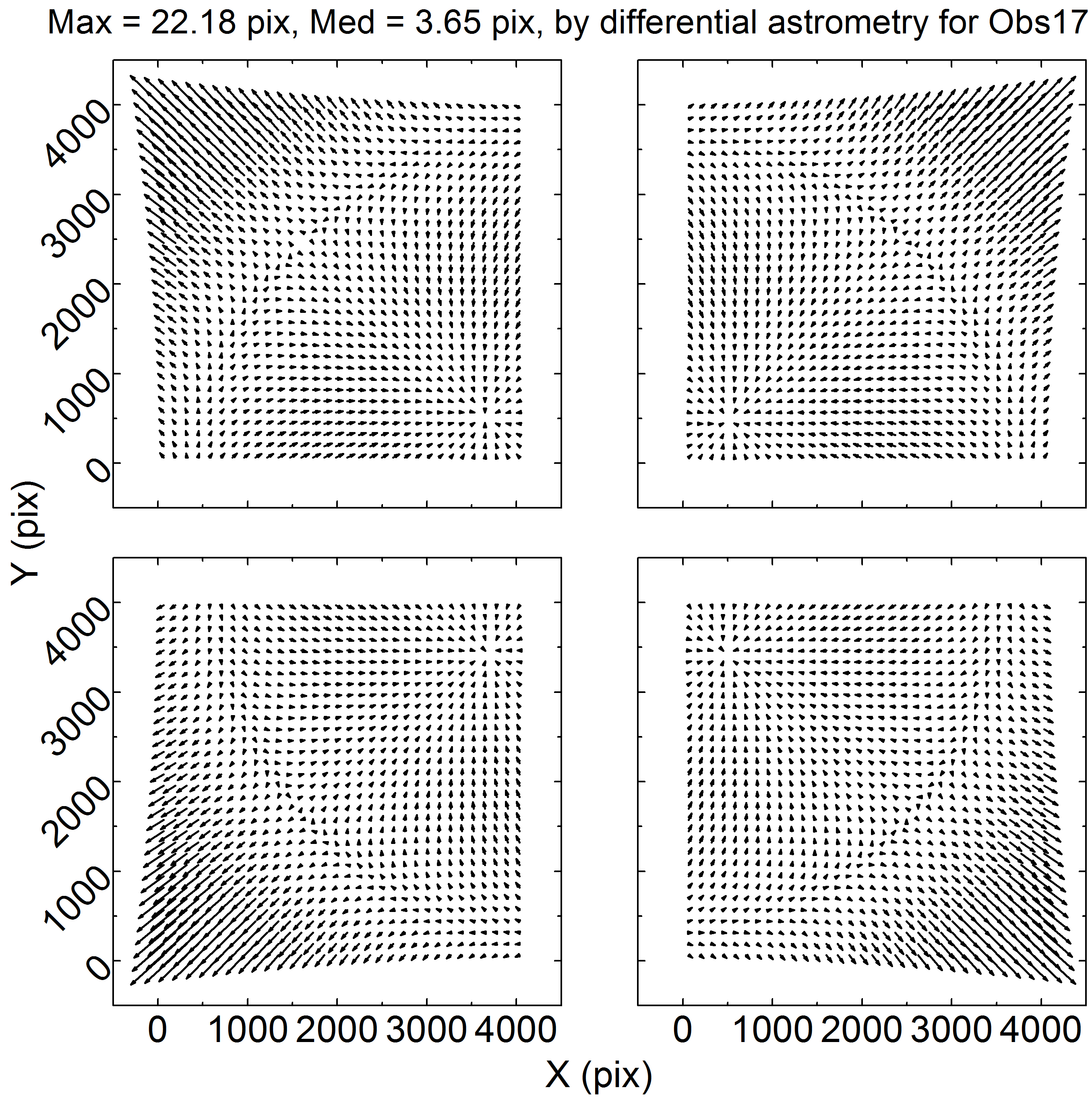}
    \end{minipage}
    \vspace{0in}
\caption{GD solution for 90Prime camera of the Bok 2.3-m telescope. The left panels show GD1 solutions derived by photographic astrometry, and the right panels show GD2 solutions derived by differential astrometry. The top panels show GD solutions derived from Obs16, and the bottom panels show those derived from Obs17. The horizontal direction corresponds to the x axis, and the vertical direction corresponds to the y axis. The vectors of the solutions are magnified by a factor of 20. Maximums and medians are shown at the top in pixels.}
\label{FigGD}
\end{figure*}

\begin{figure*}[htp]
    \begin{minipage}{\textwidth}
    \centering
        \includegraphics[width=0.48\textwidth, angle=0]{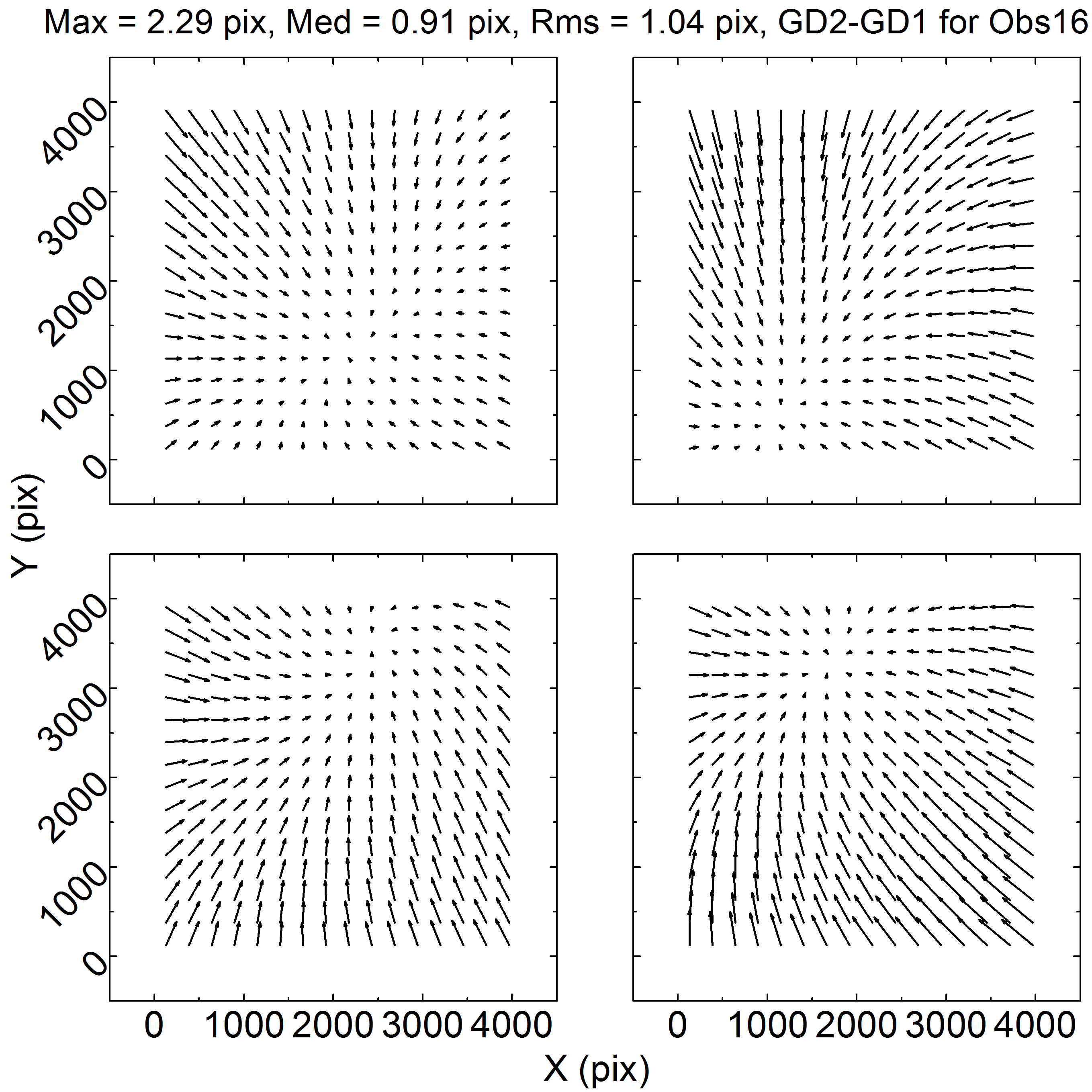}
        \includegraphics[width=0.48\textwidth, angle=0]{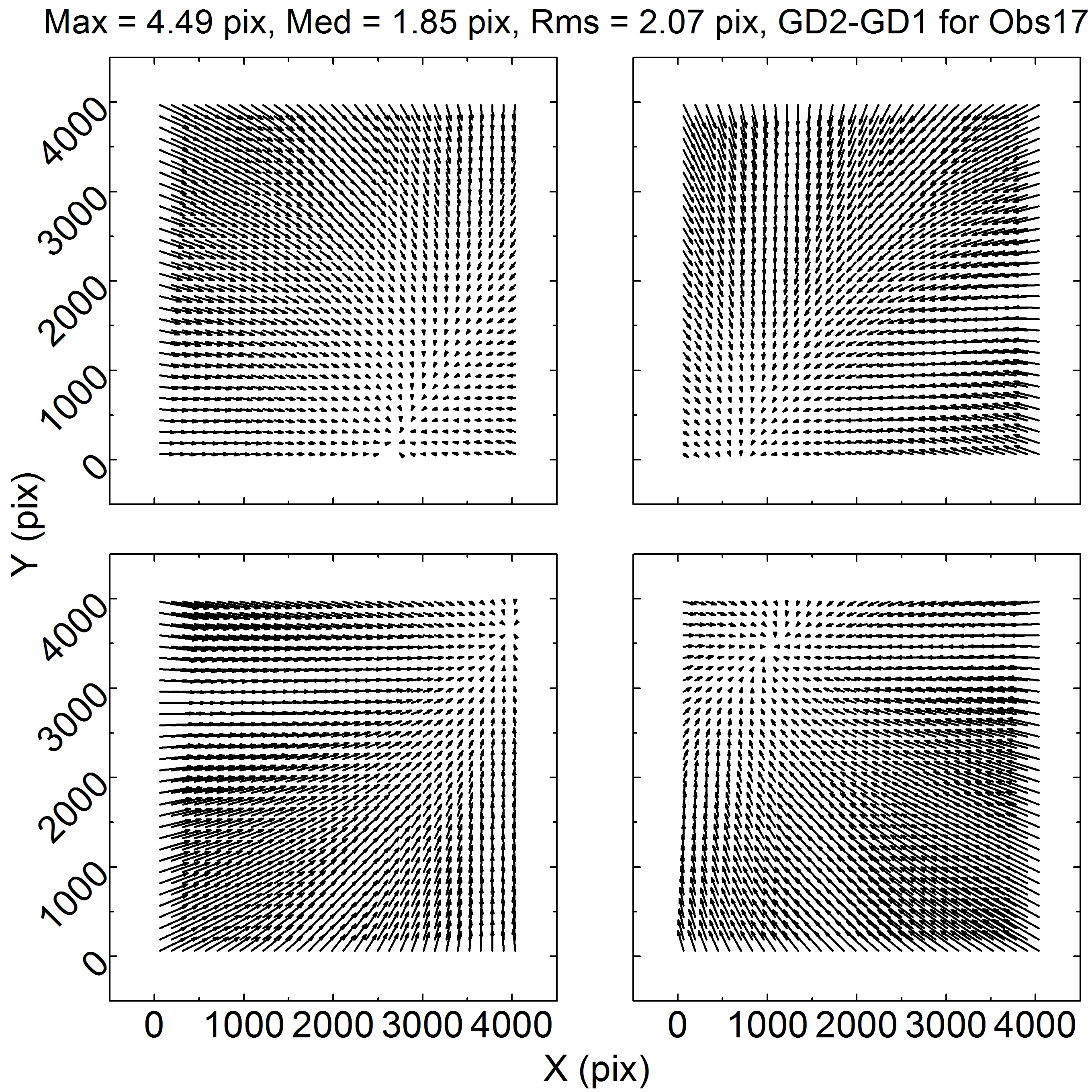}
    \end{minipage}
    \vspace{0in}
\caption{Differences between the models derived from the GD1 solution and the GD2 solution. The left panels show the differences for Obs16, and the right panels show the differences for Obs17. The horizontal direction corresponds to the x axis, and the vertical direction corresponds to the y axis. The vectors of the solutions are magnified by factors of 200 for Obs16 and 100 for Obs17, respectively. Maximums, medians, and the RMSs of the differences are shown at the top in pixels. We note that RMS=$\sqrt{\frac{\sum_{i=1}^n((x_{GD2}-x_{GD1})^2+(y_{GD2}-y_{GD1})^2)}{n-1}}$.}
\label{FigGDSub}
\end{figure*}

The derived GD2 solutions are shown in the right panel of Fig.~\ref{FigGD}, and they are similar to the GD1 solutions. A detailed comparison between them~(GD2-GD1) for the respective chip is shown in Fig.~\ref{FigGDSub}. The median differences between the two GD solutions achieve a considerable  level~(0.91 pixels for Obs16 and 1.85 pixels for Obs17). As shown in Appendix A, the cubic terms of the two GD solutions are similar, and larger differences are found in quadratic terms and linear terms. Similar discrepancies between the two GD solutions are also found in our previous paper~\citep{Zheng2021}. Although \cite{Anderson2003} stated that the observations at the same orientation are unable to constrain the linear distortion terms, we think there are also some quadratic distortion terms in the constructed master frame. As a result, there might be a degeneracy between low-order terms (order$\leq$2) in the fitted distortion polynomial and the translation in the conformal transformation. By introducing observations at various orientations, the discrepancy between the two GD solutions will become much smaller~\citep{Zheng2021}. Unlike many ground-based telescopes with rigid mounting of the telescope itself and of its detectors, the CSST and other space telescopes are able to obtain this type of observations.

\subsection{Accuracy of the GD solution}

To evaluate the accuracy of the GD solution, the GD-corrected positions were transformed to the DFF's positions. The difference between the transformed positions ($x_i^{T},y_i^{T}$) and the distortion-free positions ($\tilde{x},\tilde{y}$) directly quantifies how close we are to reaching the DFF. For example, the standard deviation~($\sigma_x$) of ($x_i^{T}-\tilde{x}$) was used to estimate the measurement precision in $x$ as follows:
\begin{equation}
\begin{aligned}
  \sigma_x = \sqrt{\frac{\sum_{i=1}^n(x_i^{T}-\tilde{x})^2}{n-1}} \\
  \nonumber
\end{aligned},
\end{equation}
and the precisions of other quantities stated in this paper are similar to the above equation.

A six-parameter linear model was used to transform GD-corrected positions to DFF positions. This is because it corrects not only the changes in linear terms of GD, which are easily changed by variations in the telescope's flexure, but also a large amount of the differential atmospheric refraction~(DAR, discussed in detail in the next sub-section) in the FOV for differential astrometry.

\begin{figure*}[htp]
\centering
\includegraphics[width=0.9\textwidth]{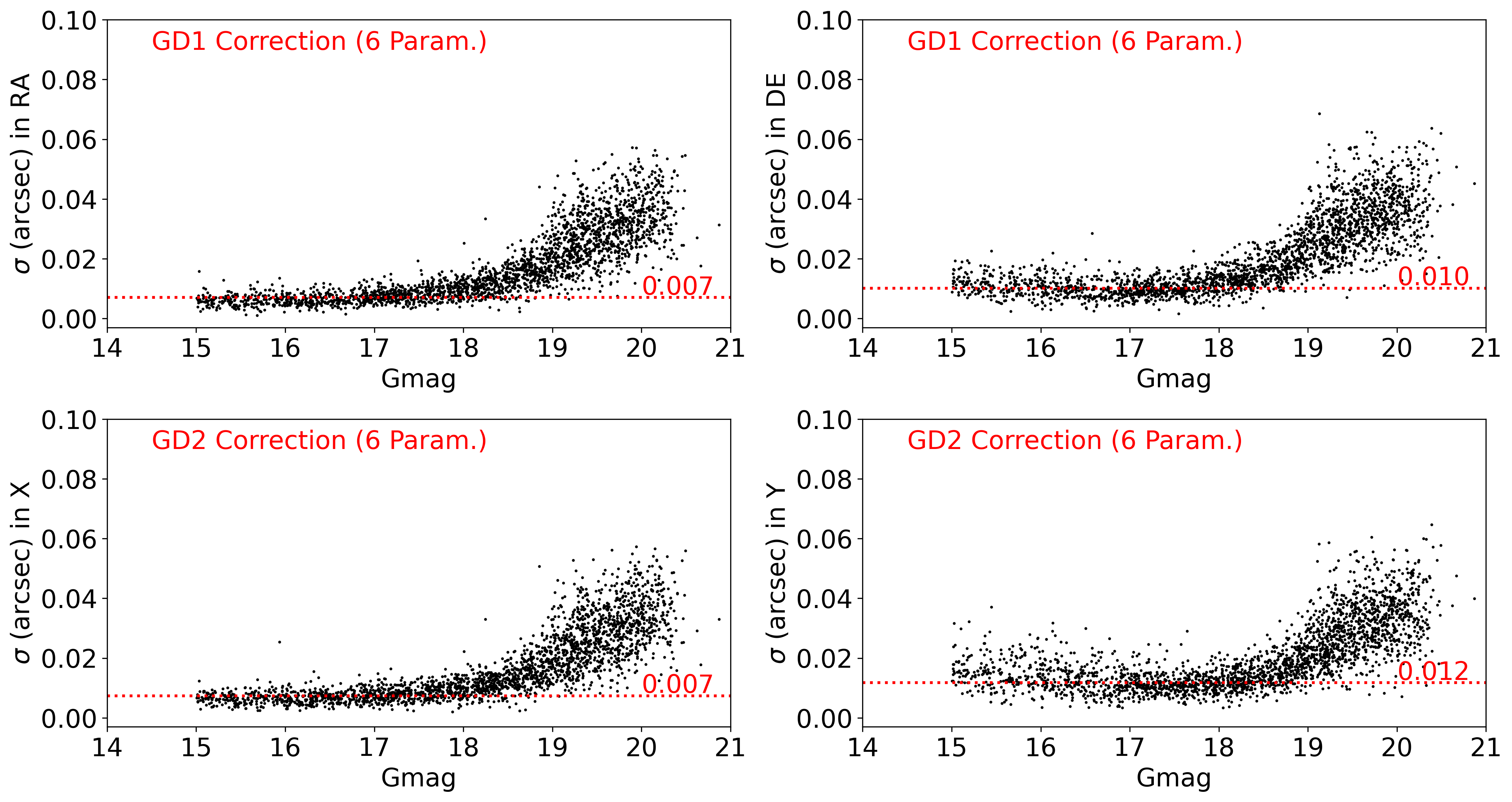}
\caption{Measurement precision for Obs16 after GD correction. Top: Standard deviation~($\sigma$) of the transformation residuals after
GD1 correction as a function of Gmag by a six-parameter linear transformation to the standard coordinate. The left shows the $RA$ direction and the right shows the $DE$ direction. Bottom: Standard deviation of the transformation residuals after
GD2 correction as a function of Gmag by a six-parameter linear transformation to the master frame. The left shows the $X$ direction and the right shows the $Y$ direction. The red dashed line marks the median ($\sigma$) for the stars brighter than 18 Gmag.}
\label{FigError16}
\end{figure*}

\begin{figure*}[htp]
\centering
\includegraphics[width=0.9\textwidth]{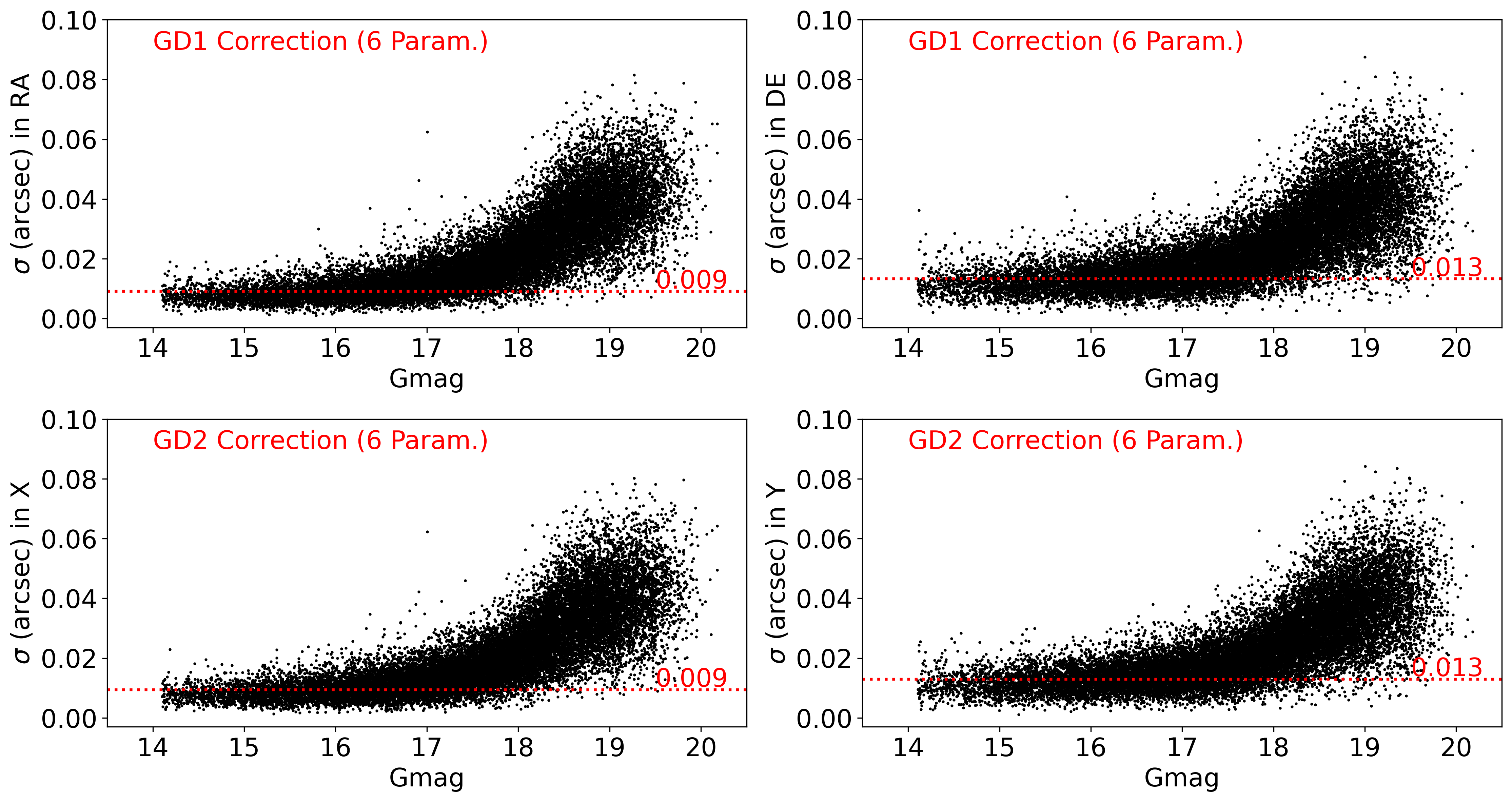}
\caption{Measurement precision for Obs17 after GD correction. Top: Standard deviation~($\sigma$) of the transformation residuals after
GD1 correction as a function of Gmag by a four-parameter linear transformation to the standard coordinate. The left shows the $RA$ direction and the right shows the $DE$ direction. Bottom: Standard deviation of the transformation residuals after GD2 correction as a function of Gmag by a six-parameter linear transformation to the master frame. The left shows the $X$ direction and the right shows the $Y$ direction. The red dashed line marks the median ($\sigma$) for the stars brighter than 17 Gmag. It should be noted that the GD solution of the top panel is derived by means of photographic astrometry and the GD solution of the bottom panel is derived by means of differential astrometry.}
\label{FigError17}
\end{figure*}

The measurement precisions after GD correction for Obs16 and Obs17 are shown on the bottom panel of Fig.~\ref{FigError16} and Fig.~\ref{FigError17}. As the $x$ axis is nearly parallel to the $RA$ direction and the $y$ axis is nearly parallel to the $DEC$ direction, to facilitate the comparison, the results of differential astrometry are converted to arcseconds with an average pixel scale of $0\farcs453$ pix$^{\text{-}1}$.

For Obs16, the precision of differential astrometry is slightly inferior to photographic astrometry in the $Y$ direction or the approximate $DEC$ direction, although it just tells us how accurately we can expect to register the relative position of a star among different dithered frames. For Obs17, the precisions of the two types astrometry reach a comparable level. The precision in the $RA$~(or $X$) direction is smaller than that in the $DEC$~(or $Y$) direction. The situation is similar to~that of \cite{Wang2019}~(Fig. 20). Thanks to the excluding criteria, fewer abnormal values above the average level are found in our results, especially for stars fainter than 17 Gmag.

\begin{figure}[htp]
\centering
\includegraphics[width=\columnwidth]{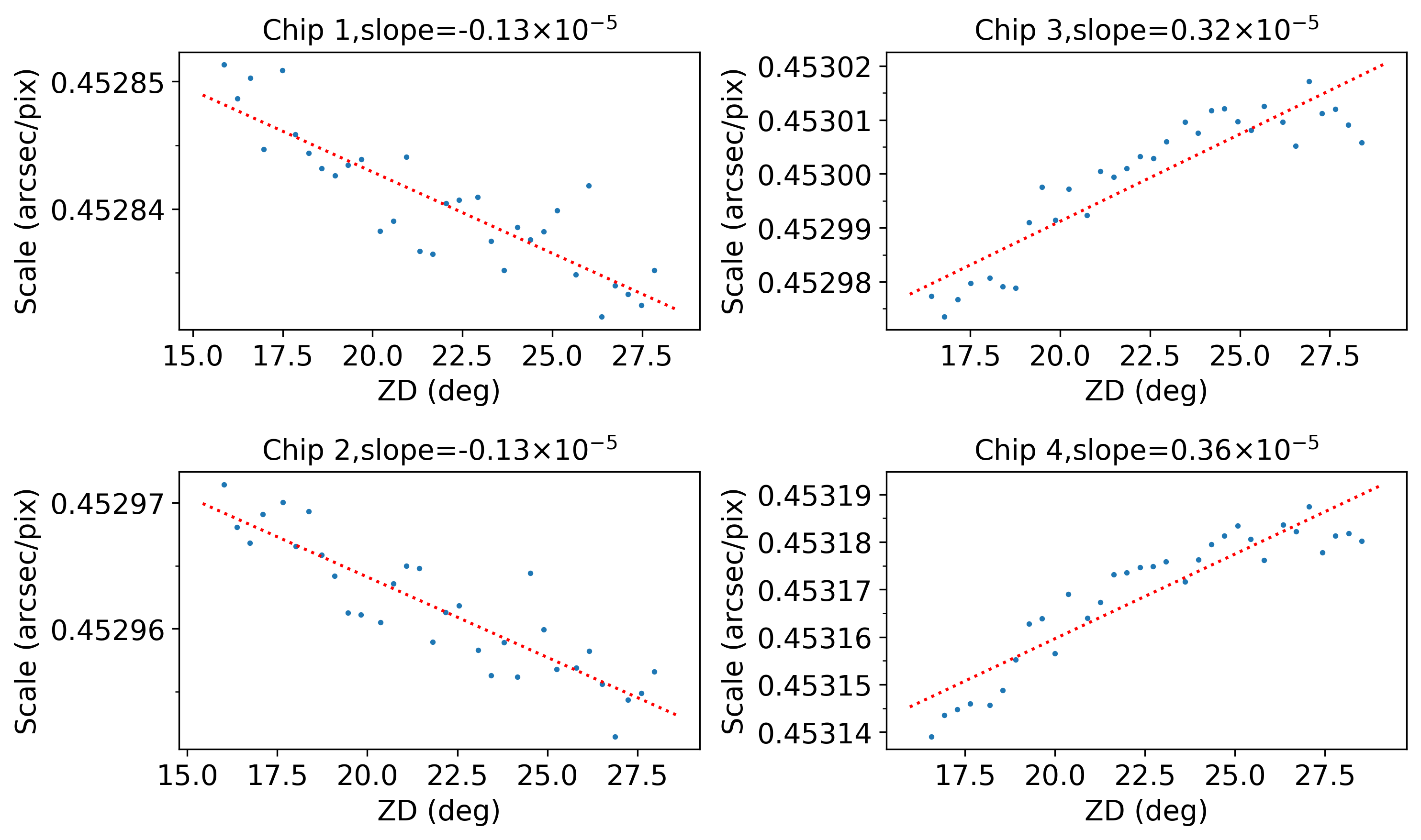}
\caption{Variations of the physical pixel scale for Obs16 against the zenith distance. The variations can be estimated by the slope of a fitted straight line to the data.}
\label{FigScale16}
\end{figure}

\begin{figure}[htp]
\centering
\includegraphics[width=\columnwidth]{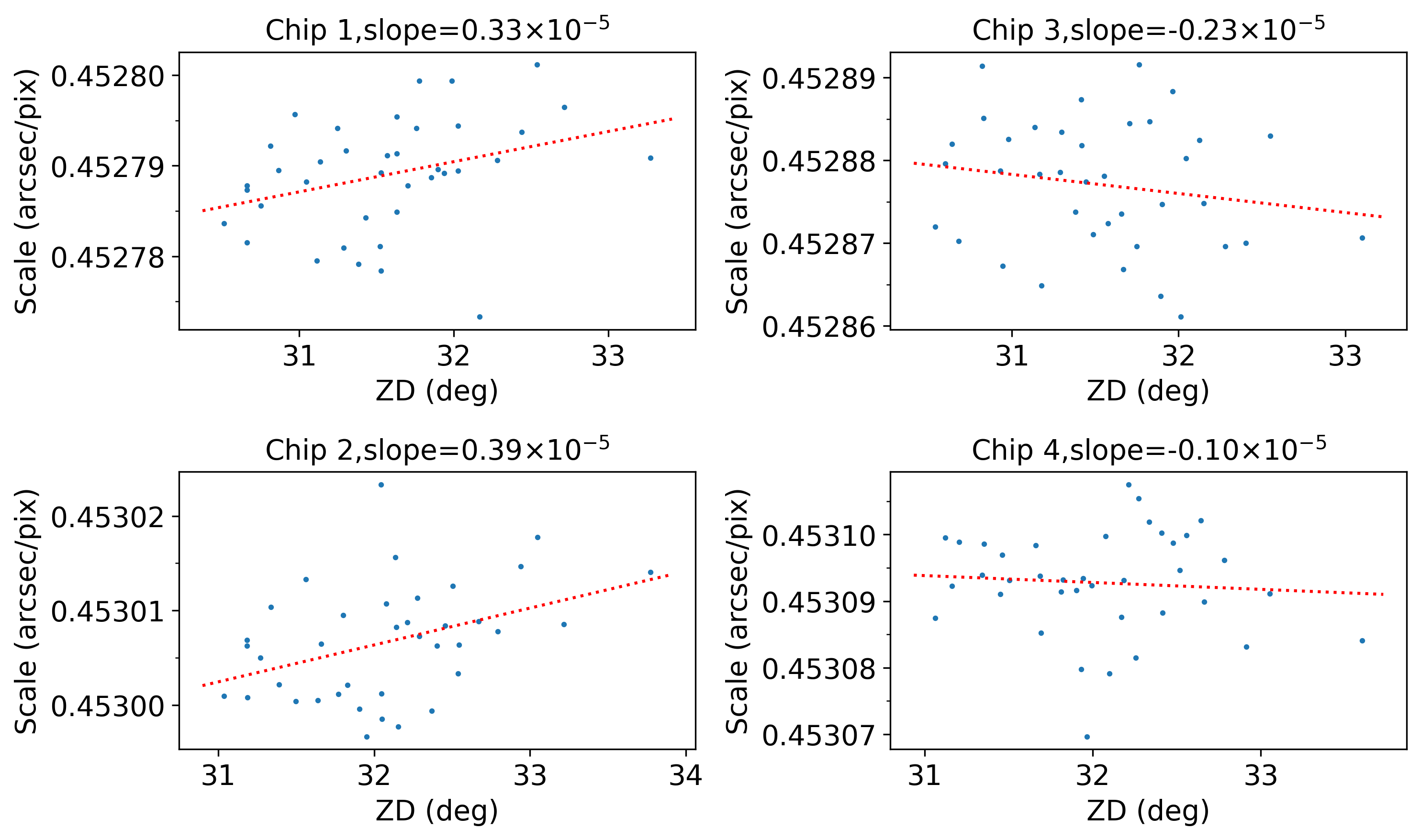}
\caption{Variations of the physical pixel scale for Obs17 against the zenith distance. The variations can be estimated by the slope of a fitted straight line to the data.}
\label{FigScale17}
\end{figure}

\begin{figure}[htp]
\centering
\includegraphics[width=\columnwidth]{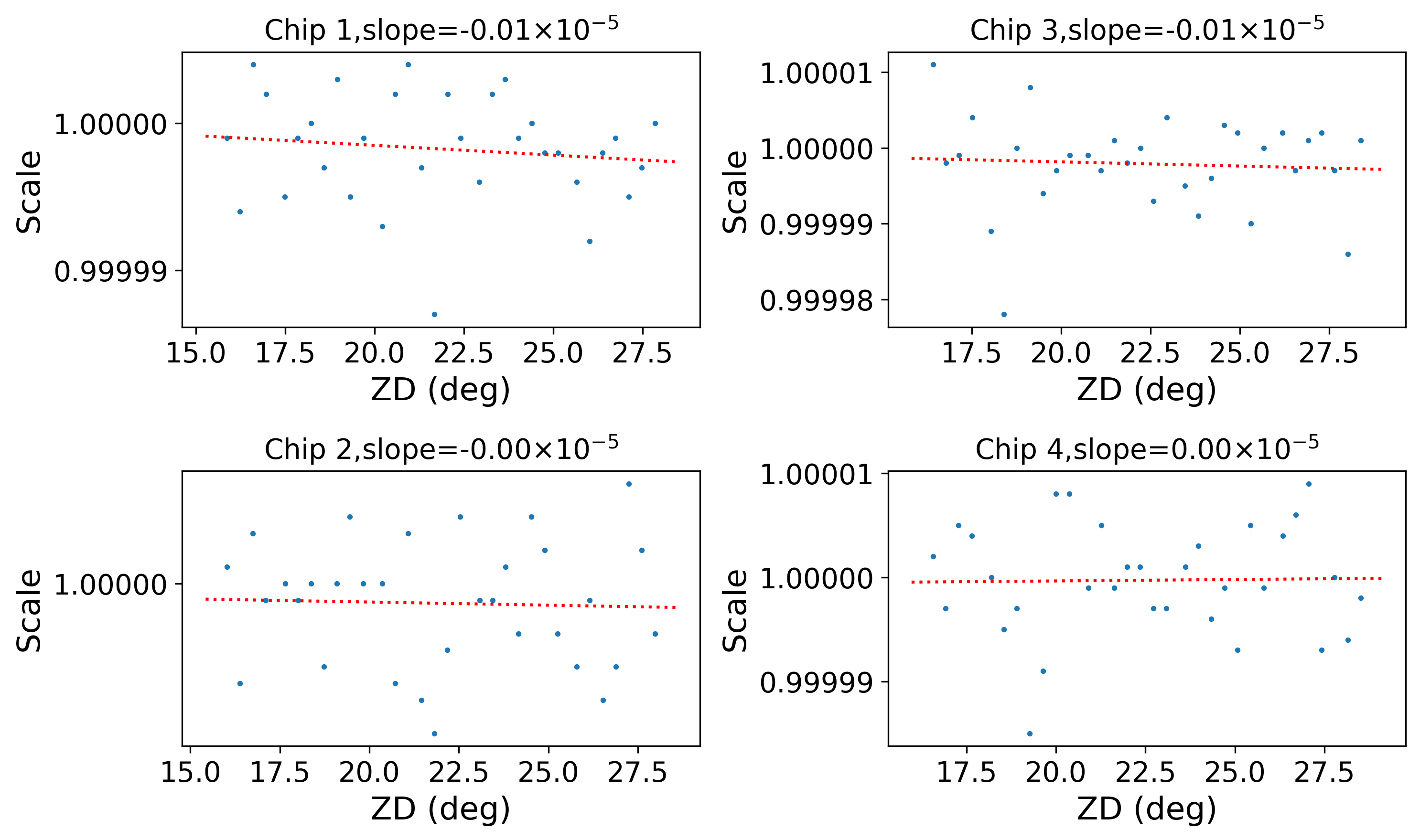}
\caption{Variations of the scale relative to the master frame for Obs16 against the zenith distance. The variations can be estimated by the slope of a fitted straight line to the data.}
\label{FigScaleM16}
\end{figure}

\begin{figure}[htp]
\centering
\includegraphics[width=\columnwidth]{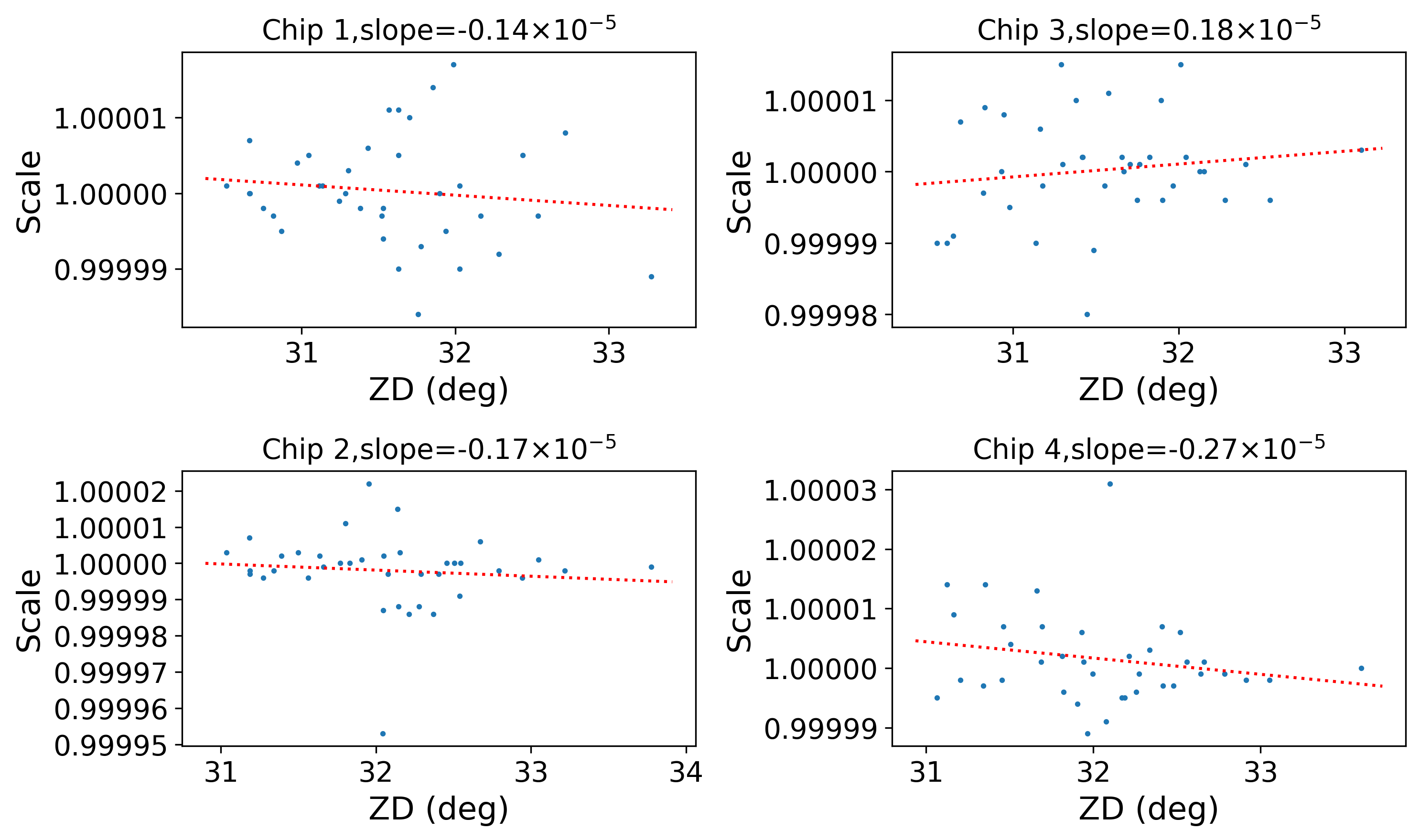}
\caption{Variations of the scale relative to the master frame for Obs17 against the zenith distance. The variations can be estimated by the slope of a fitted straight line to the data.}
\label{FigScaleM17}
\end{figure}

As \cite{Anderson2003} indicated, there is an interaction between scale and distortion. Therefore, we investigate if the scale varies during the observation. The pixel scale against the zenith distance is shown in Fig.~\ref{FigScale16} for Obs16. The pixel scale against the zenith distance of Obs17 is shown in Fig.~\ref{FigScale17}. For the two epochs, the pixel scales change very little during the observation; although, a close dependence of pixel scale on zenith distance can be found for Obs16. We thus conclude that the GD effects are stable during the observation.

We also investigate the variations of the scale relative to the master frame, and the results are shown in Fig.~\ref{FigScaleM16} and Fig.~\ref{FigScaleM17}. A similar conclusion is reached. The stability of the scale also confirmed that much of the projection effects have been reduced since the local master frames have nearly the same scale as the root observations.

\subsection{Estimation of the DAR of the observations for differential astrometry}
Since the 90Prime camera has such a large FOV, the DAR on the observations should be considered when using differential astrometry. The DAR for two sources in arcsec with observed zenith distance $z$ and zenith separation $\Delta z_{0}$, is given approximately by \citet{Gubler1998}:
\begin{equation}
\begin{aligned}
{\rm DAR} &= \frac{(1+\tan^2z)(A+3B\tan^2z)\Delta z_{0}}{206265}  \\
\nonumber
\end{aligned},
\end{equation}
where $A$ and $B$ depend on the meteorological conditions~(atmospheric pressure, ambient temperature and relative humidity can be found from the historic data of the website for the site weather of Kitt Peak National Observatory\footnote{\url{http://www-kpno.kpno.noao.edu/Info/Mtn\_Weather/}}) and the wavelength of the filter.

To estimate the DAR on the observations, we used iauRefco, a C routine from SOFA library to compute $A$ and $B$. The adopted value of the input parameters for iauRefco function are listed in Table~\ref{Tab3}. $\Delta z_{0}$ is set as the largest separation between a pair of stars as the FOV of each chip~($\approx$0.5${\degr}$). Finally, we find that the DAR is mainly a linear trend for the observations, since the largest residual after using a line to fit the DAR just reaches the minimum positional error; that is, 0.01 pixels~($\sim$5 mas). Therefore, the DAR is considered negligible when using a six-parameter linear transformation for differential astrometry.

\begin{table}
\caption{Specifications of the input parameter for iauRefco.}
\centering
\begin{tabular}{@{}ccccccc@{}}
\hline\hline
Date & PHPA & TC & RH  &  WL  \\
&millibar & ${\degr}$C &  & $\mu$m\\
\hline
2016.1.17& 778 & 5.7 & 50$\%$ &  0.6412 \\
2017.3.5 & 805 & 9.2 & 40$\%$ &  0.4776 \\
\hline
\end{tabular}
\label{Tab3}
\end{table}

\section{The determination for relative positions between the CCD chips}

\subsection{Solution to the geometry of CCD mosaic chips -- Methodology}

In this section, we interpret the procedure of determination of the geometry of CCD
mosaic chips by taking advantage of the derived GD solution. The procedure is mainly explained in the way of photographic astrometry as follows.

For the same pointing, the observations taken with four chips share the same standard coordinate derived from Gaia EDR3. We adopted a six-parameter linear transformation to relate the standard coordinate~$(\xi,\eta)$ to the pixel position~$(x,y)$ after GD correction~(GDC) for the observations, since it can absorb not only the linear distortion which is vulnerable to the gravitational flexure effect, but also the DAR in differential astrometry. The pixel position~$(x,y)$ after GDC on each CCD chip can be transformed into the standard coordinate~$(\xi,\eta)$.
In turn, the standard coordinate can be transformed into the distortion-free pixel position of any chip. Furthermore, due to the distortion effect, its actual or physical pixel position on the chip can be estimated by implementing reverse GDC to the distortion-free position. Throughout the process, the standard coordinate serves as a bridge to enable the pixel positions of one chip to be transformed into the pixel coordinate of another chip.

\begin{figure}[htp]
\centering
\includegraphics[width=\columnwidth, angle=0]{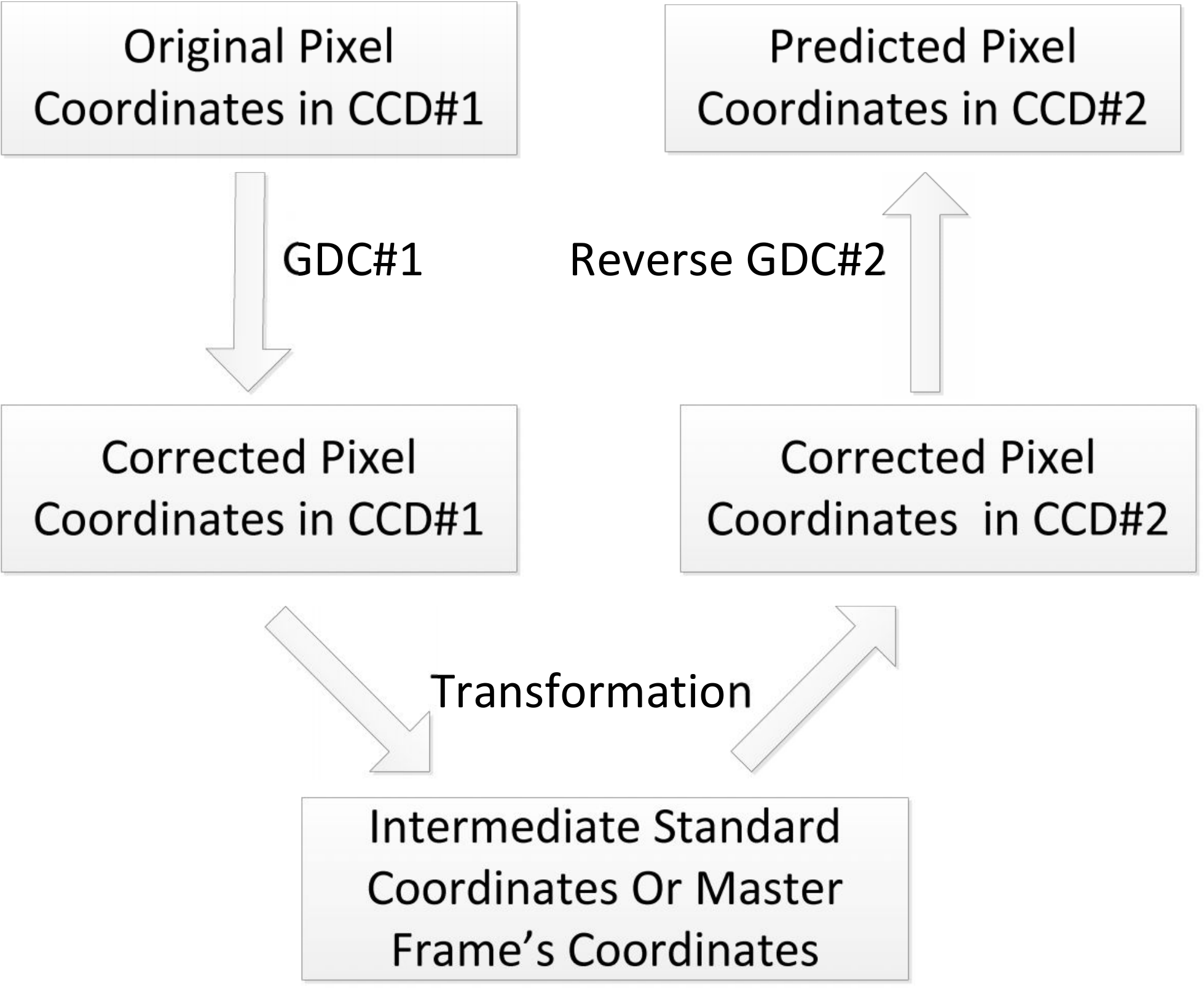}
\caption{Flow chart of transforming edge L$_{3\text{-}4}$ in CCD\#1 into the pixel coordinate of the reference chip CCD\#2 through photographic astrometry or differential astrometry. The key distinctions between the two methods are the 'bridge' linking two adjacent CCD chips, which are the standard coordinates, or the master frame, respectively.}
\label{Fig3}
\end{figure}

Specifically, we calculated the relative positions of two adjacent edges between chips to estimate the geometry of CCD mosaic chips. To further illustrate the process, we first give a layout of the CCD mosaic chips and designate a number for each corner of the chips, as shown in Fig.~\ref{Fig2}. L$_{m\text{-}n}$ would represent a horizontal or vertical edge from Corner m to Corner n~(all edges are in red in~Fig.~\ref{Fig2}). Any edge of a chip can be transformed to the physical pixel coordinate of the adjacent reference chip in order to estimate the relative positions between the chips. In the following, we take the transformation of L$_{3\text{-}4}$ into the pixel coordinate of CCD\#2 as an example. The whole procedure is shown in~Fig.~\ref{Fig3}.

As mentioned above, L$_{3\text{-}4}$ in CCD\#1 after GDC of CCD\#1~(GDC\#1 for short) can be transformed into the standard coordinate for each frame. Similarly, L$_{3\text{-}4,}$ in standard coordinates, can be transformed into CCD\#2's distortion-free pixel coordinate, depending on the transformation of CCD\#2 into the standard coordinate. Finally, considering the actual GD effect of CCD\#2, we adopted a reverse GDC\#2 for L$_{3\text{-}4}$ to estimate its physical~(distorted) positions in CCD\#2's pixel coordinate, as described below.

As seen from Fig.~\ref{FigGD}, the GD effects of the adjacent regions of any two chips are relatively small, while the GD effects reach the maximum near the four corners of the FOV. So, firstly, we roughly estimated the distorted positions. That is, we used the undistorted position ($x_0, y_0$) as the input for the GD model of CCD\#2, to compute GD effects. The distorted positions ($x_d, y_d$) can
be derived with the undistorted positions and the corresponding GD effects. Then, the distorted positions were taken as the new input for the GD model to derive new undistorted pixel positions ($\tilde{x}_0, \tilde{y}_0$). The differences between the original undistorted positions ($x_0, y_0$) and the new undistorted pixel positions ($\tilde{x}_*, \tilde{y}_*$) were used to adjust the distorted positions, leading to much smaller differences. Finally, the iteration stopped until the differences were less than 0.001 pixels. Based on the estimated distorted pixel positions, the relative position between CCD\#1 and CCD\#2 can be solved analytically.

It should be noted that the GD solution for each chip has the effective area covering the size of the chip, that is 4032${\times}$4096, therefore one should avoid transforming an edge too far from the reference chip, which would introduce excess extrapolation errors when performing a reverse GDC. Therefore, we choose L$_{3\text{-}4}$ rather than L$_{1\text{-}2}$ to be transformed into CCD\#2.

However, if there are a lot of stars fainter than the Gaia faint limit~($\sim$21 Gmag), such as in the observations of the CSST, it is more effective to implement the procedure directly based on stars' pixel positions. Accordingly, we provide an alternative procedure by means of differential astrometry to determine relative positions between CCD chips. In a similar manner, every pixel position of a chip can be transformed to the pixel coordinate of another chip, by using the local master frame and the GD solution~(Fig.~\ref{Fig3}). Therefore, it is not repeated here.

\section{Results and discussions}
\begin{figure}
\centering

\includegraphics[width=0.5\textwidth, angle=0]{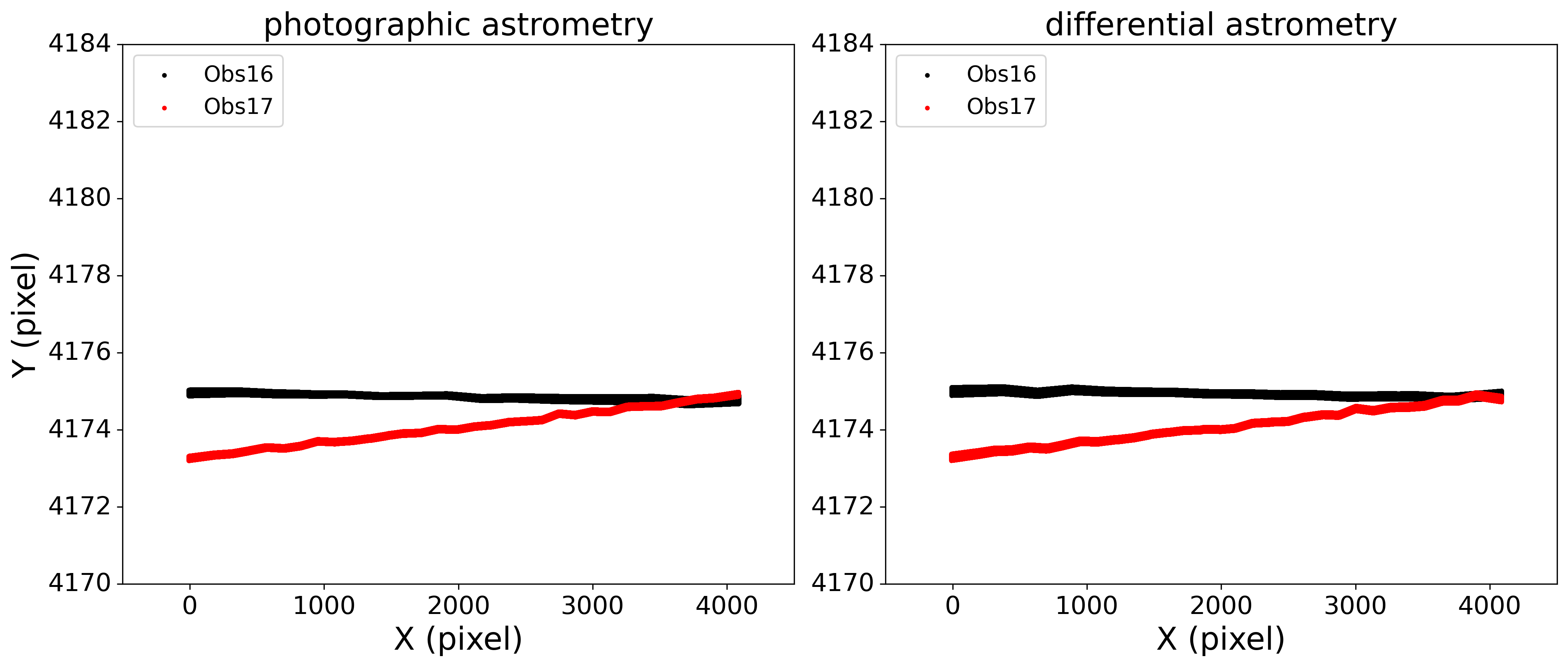}

\caption{Transformed positions of L$_{3\text{-}4}$ in the pixel coordinate of CCD\#2. The left is derived from photographic astrometry and the right is derived from differential astrometry. The black line shows the results of Obs16 and the red line shows the results of Obs17.}
\label{Fig1Ref2H}
\end{figure}

\begin{figure}
\centering

\includegraphics[width=0.5\textwidth, angle=0]{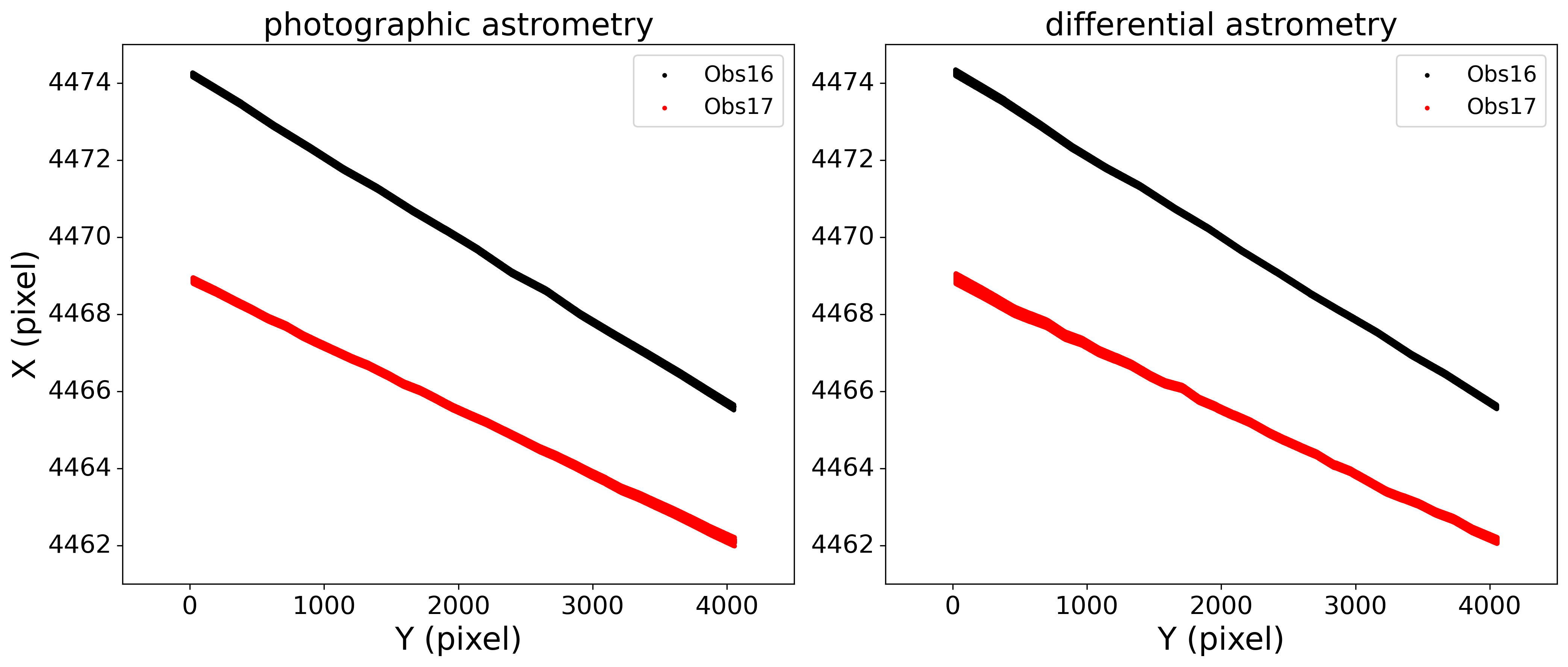}

\caption{Similar to Fig.~\ref{Fig1Ref2H}, but for the transformed positions of L$_{13\text{-}15}$ in the pixel coordinate of CCD\#2.}
\label{Fig4Ref2V}
\end{figure}

\begin{figure}
\centering

\includegraphics[width=0.5\textwidth, angle=0]{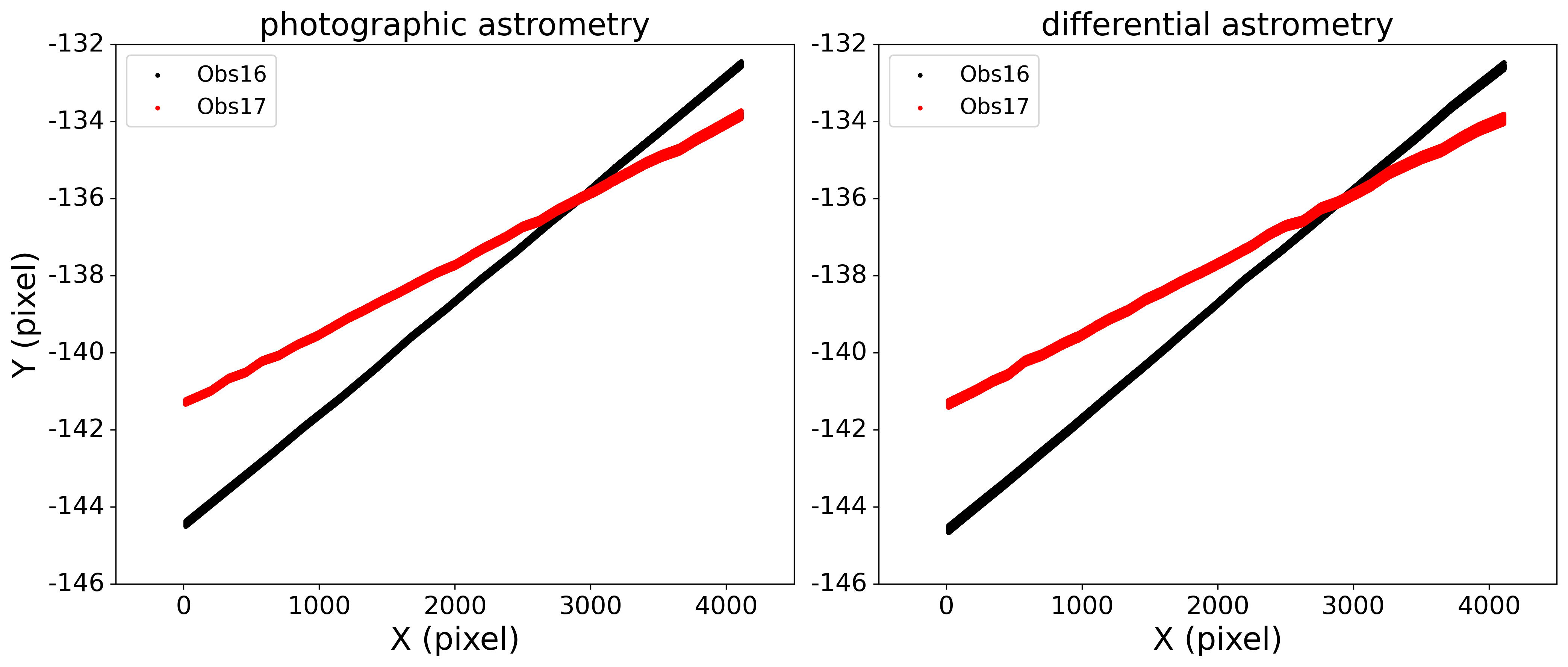}

\caption{Similar to Fig.~\ref{Fig1Ref2H}, but for the transformed positions of L$_{13\text{-}14}$ in the pixel coordinate of CCD\#3.}
\label{Fig4Ref3H}
\end{figure}

\begin{figure}
\centering
\includegraphics[width=0.5\textwidth, angle=0]{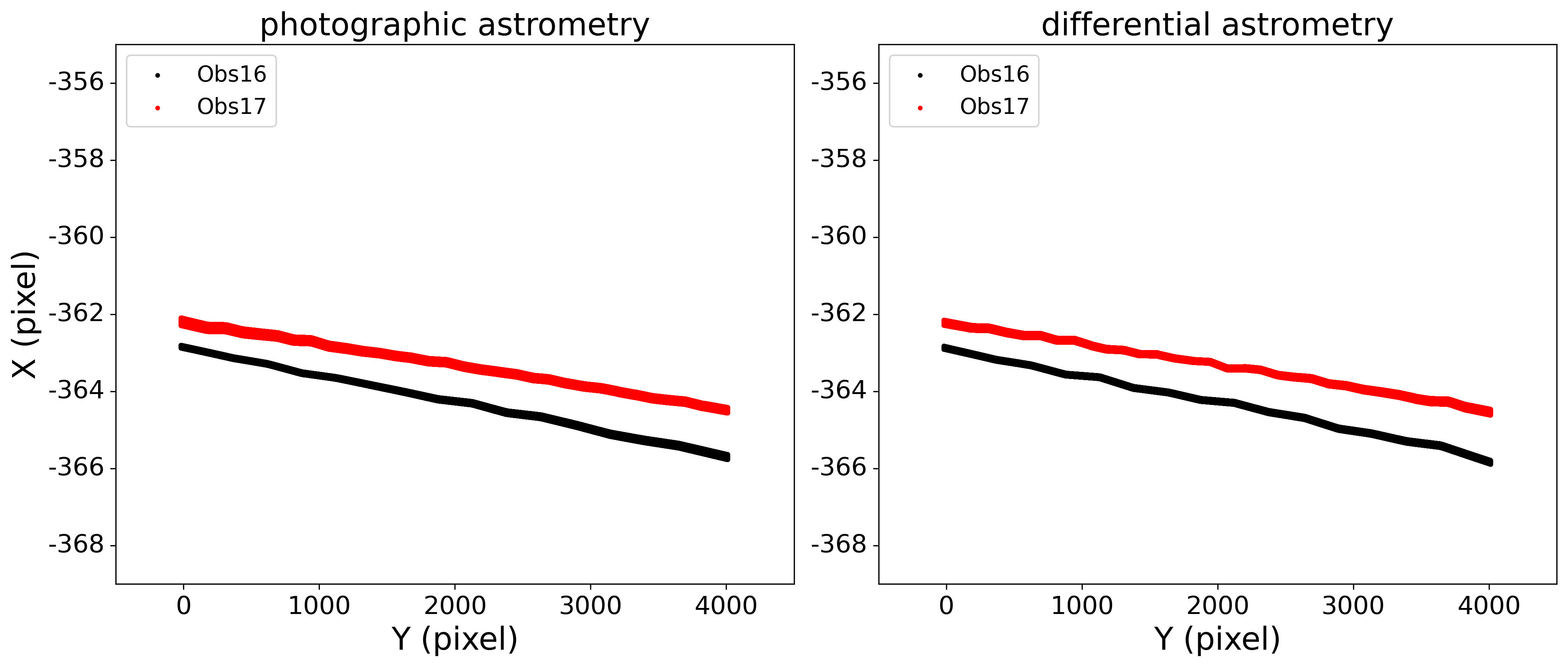}
\caption{Similar to Fig.~\ref{Fig1Ref2H}, but for the transformed positions of L$_{2\text{-}4}$ in the pixel coordinate of CCD\#3.}
\label{Fig1Ref3V}
\end{figure}

\begin{figure}
\centering

\includegraphics[width=0.5\textwidth, angle=0]{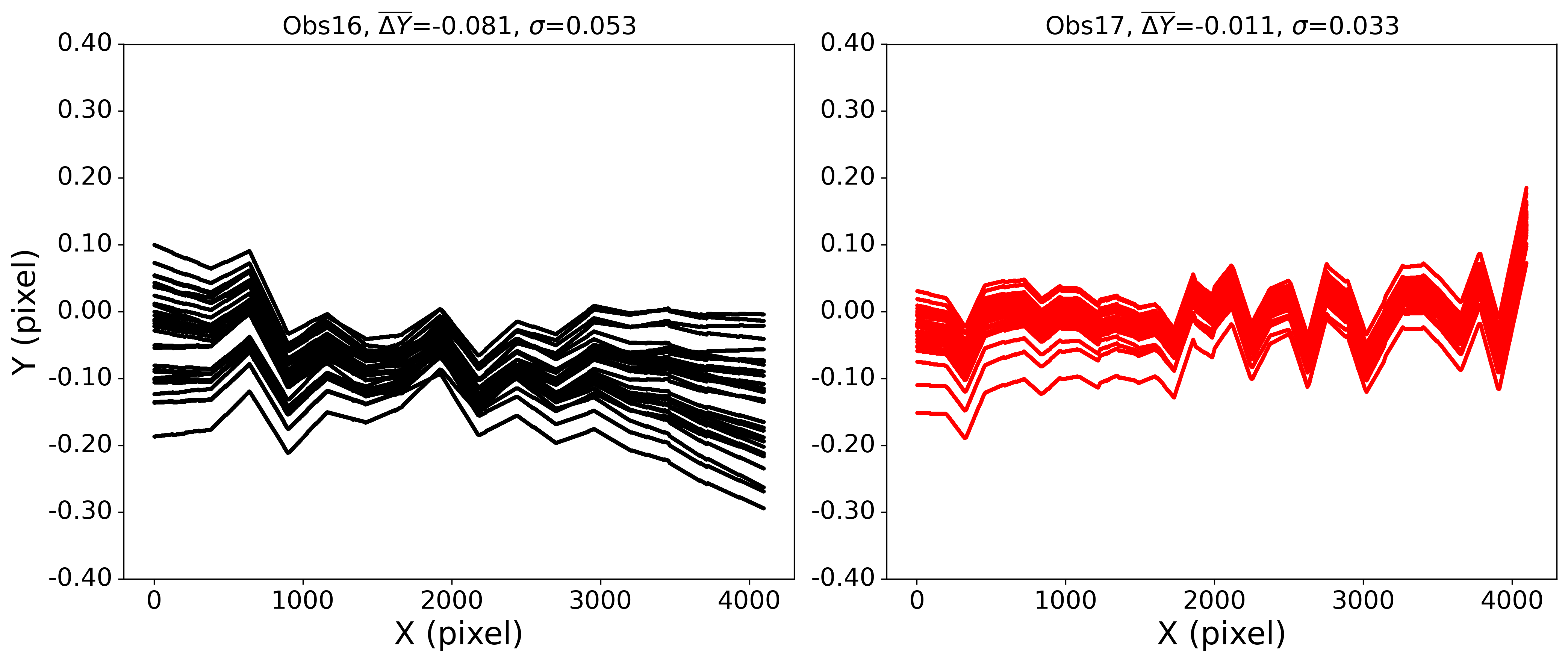}

\caption{Differences between the transformed positions of L$_{3\text{-}4}$ in the pixel coordinate of CCD\#2 derived from the two types of astrometry. Each line corresponds to a single observation. The left is for Obs16 and the right is for Obs17. The averages and the standard deviation~($\sigma$) of the differences are shown at the top in pixels. }
\label{Fig1Ref2H_sub}
\end{figure}

\begin{figure}
\centering

\includegraphics[width=0.5\textwidth, angle=0]{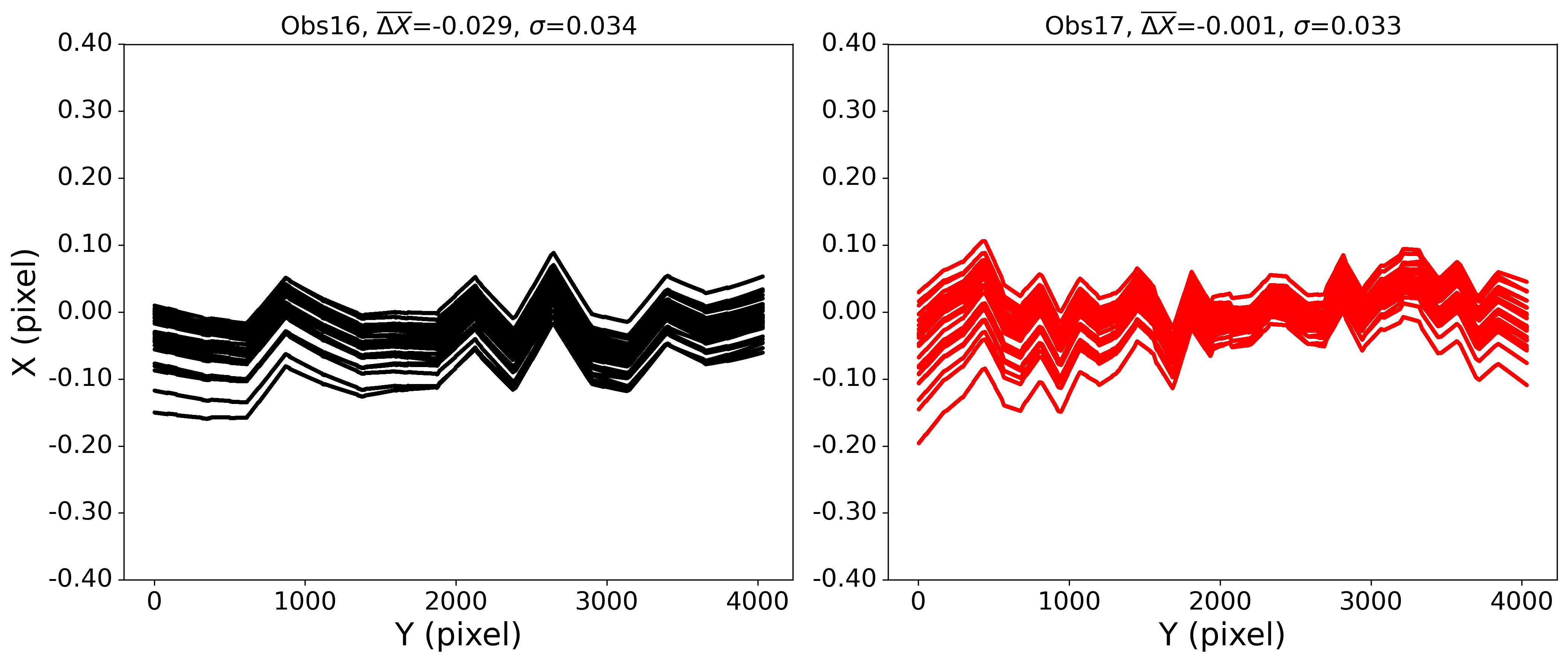}

\caption{Similar to Fig.~\ref{Fig1Ref2H_sub}, but for the transformed positions of L$_{13\text{-}15}$ in the pixel coordinate of CCD\#2. Each line corresponds to a single observation.}
\label{Fig4Ref2V_sub}
\end{figure}

\begin{figure}
\centering

\includegraphics[width=0.5\textwidth, angle=0]{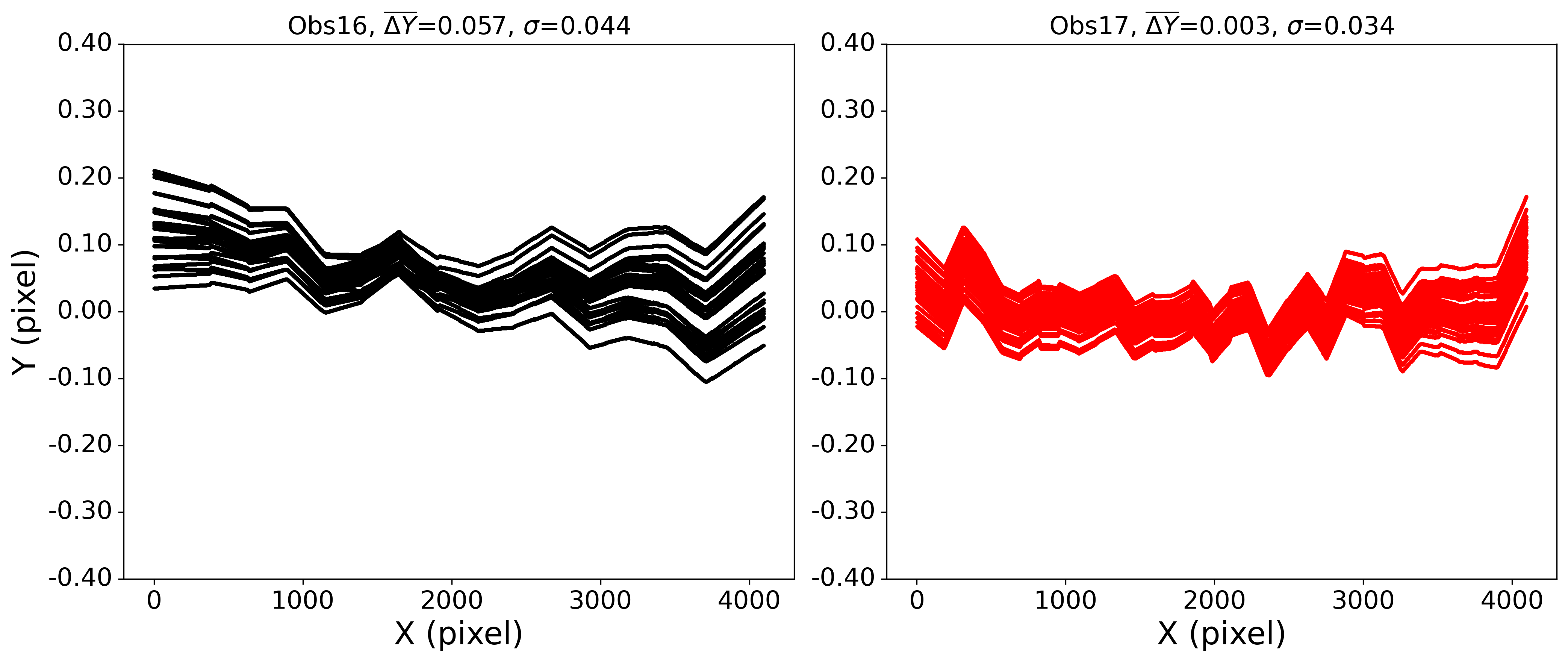}

\caption{Similar to Fig.~\ref{Fig1Ref2H_sub}, but for the transformed positions of L$_{13\text{-}14}$ in the pixel coordinate of CCD\#3. Each line corresponds to a single observation.}
\label{Fig4Ref3H_sub}
\end{figure}

\begin{figure}
\centering
\includegraphics[width=0.5\textwidth, angle=0]{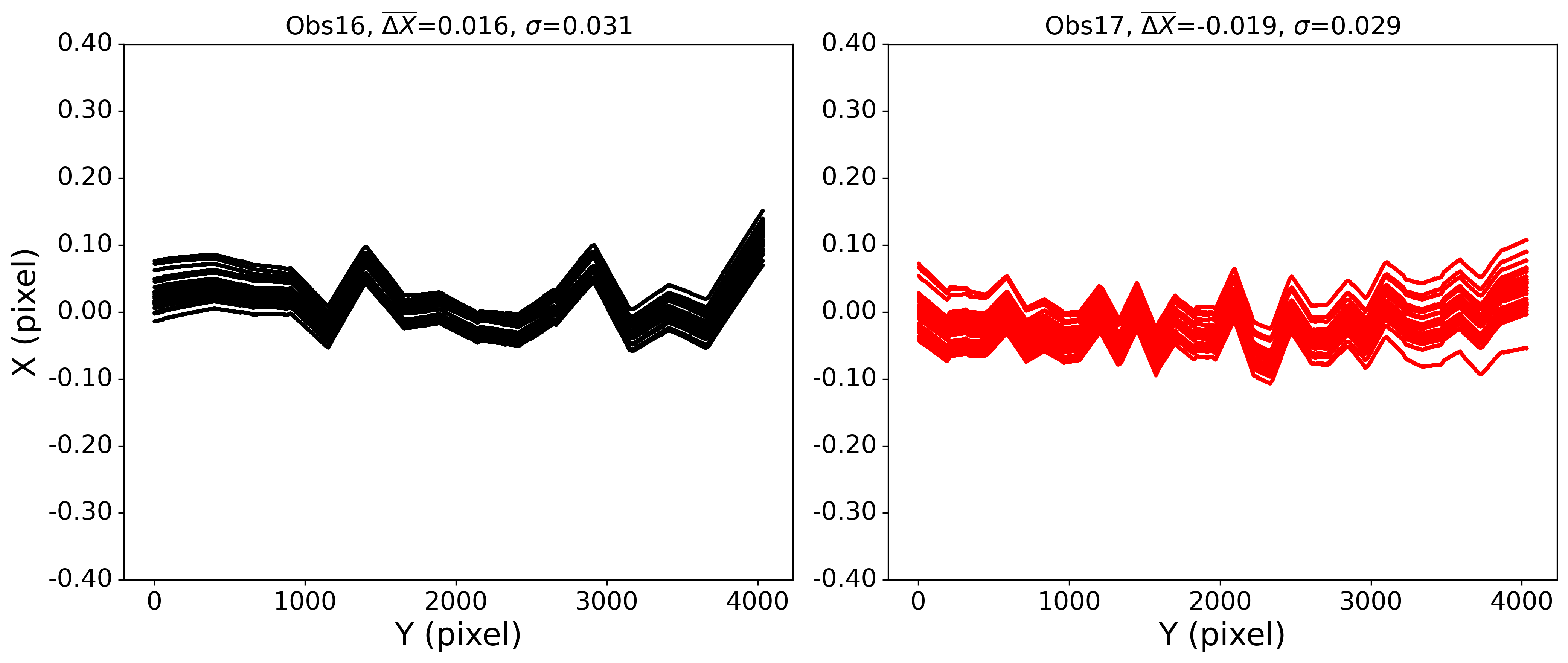}
\caption{Similar to Fig.~\ref{Fig1Ref2H_sub}, but for the transformed positions of L$_{2\text{-}4}$ in the pixel coordinate of CCD\#3. Each line corresponds to a single observation.}
\label{Fig1Ref3V_sub}
\end{figure}

In this section, we adopted a six-parameter linear transformation to transform the pixels of L$_{3\text{-}4}$ and L$_{13\text{-}15}$ into the pixel coordinate of CCD\#2 and to transform L$_{13\text{-}14}$ and L$_{2\text{-}4}$ into CCD\#3. The results of the two epochs are plotted from Fig.~\ref{Fig1Ref2H} to Fig.~\ref{Fig1Ref3V}, respectively. It is clear that the adjacent edges between any two chips are not parallel. The results of photographic astrometry and differential astrometry are similar to each other, while the plottings of \cite{Wang2019}~(Fig. 21 and Fig. 22) are clearly different. It confirms that the adopted tangential points in this paper meet the actual situation and do not introduce extra errors.

To further investigate the consistency between photographic astrometry and differential astrometry, we performed a subtraction for the relative positions derived from the two types of astrometry in the $X$ or $Y$ direction, as shown from Fig.~\ref{Fig1Ref2H_sub} to Fig.~\ref{Fig1Ref3V_sub}. For Obs16, the average of the differences between the two types of astrometry is about 0.046 pixels~($\sim$0.021 arcsec), and for Obs17, it is at much smaller level, of about 0.009 pixels~($\sim$0.004 arcsec). The average standard deviations of the differences are about 0.041 pixels~($\sim$0.019 arcsec) for Obs16 and about 0.032 pixels~($\sim$0.014 arcsec) for Obs17, respectively. We think the results of Obs17 are more credible, since there are more stars in the observations of 2017 than 2016.

Greater differences and dispersions can also be found near the marginal areas. We think the differences are mainly due to the difference between the adopted GD solution shown in Fig.~\ref{FigGDSub}, while the dispersions are mainly due to the extrapolation errors when performing reverse GDC to transform positions to the reference chip. We believe the extrapolation errors are also related to the magnitude of the GD effects in the grid elements. More extrapolation errors may be introduced if the adopted GD effects are larger. Since the GD effects of the Bok 2.3-m telescope are radial, they reach their maximum in the four corners of the FOV, as shown in Fig.~\ref{FigGD}, while the adopted GD effects for determining the relative positions between chips are much smaller.

To estimate the upper bound of the extrapolation errors, we computed the averages and the precisions of the transformed corners' position, since they are easily disturbed by the extrapolation from the GD effects. The results are shown in Table~\ref{tab4}. The averages of the differences between the results of photographic astrometry and differential astrometry are about 0.018 pixels~($\sim$0.008 arcsec) for Obs16 and about 0.026 pixels~($\sim$0.012 arcsec) for Obs17, respectively. The average precisions are about 0.049 pixels~($\sim$0.022 arcsec) for Obs16 and about 0.056 pixels~($\sim$0.025 arcsec) for Obs17, respectively.

\begin{sidewaystable}[tp]
\caption{The transformed positions of the adopted corners to the reference chips (see Fig.~\ref{Fig2}).} 
\centering 
\begin{tabular}{lcccccccccc} 
\hline\hline\noalign{\smallskip} 
 \multirow{2}*{Date}&Corner&\multicolumn{2}{c}{3}& \multicolumn{2}{c}{4}&\multicolumn{2}{c}{13}& \multicolumn{2}{c}{15} \\
 \cmidrule(r){3-4}  \cmidrule(r){5-6} \cmidrule(r){7-8} \cmidrule(r){9-10}
& Method& mean& $\sigma$ & mean& $\sigma$&  mean& $\sigma$ & mean& $\sigma$\\ [0.5ex]
\hline\noalign{\smallskip} 
\multirow{2}*{2016}&GD1&(-7.409,4174.955)&(0.020,0.043)&(4087.713,4174.804)&(0.015,0.059)&(4465.584,4050.774)&(0.031,0.043)&(4474.233,19.740)&(0.022,0.060)\\
&GD2&(-7.400,4174.983)&(0.027,0.042)&(4087.696,4174.942)&(0.018,0.029)&(4465.591,4050.840)&(0.026,0.032)&(4474.262,19.664)&(0.034,0.021)\\
\multirow{2}*{2017}&GD1&(-8.668,4173.269)&(0.028,0.034)&(4086.377,4174.928)&(0.028,0.044)&(4462.119,4054.050)&(0.023,0.031)&(4468.875,23.009)&(0.035,0.025)\\
&GD2&(-8.730,4173.295)&(0.031,0.055)&(4086.362,4174.791)&(0.029,0.046)&(4462.143,4053.977)&(0.038,0.050)&(4468.928,23.057)&(0.066,0.065)\\
\hline 
\\
\hline\hline\noalign{\smallskip} 
 \multirow{2}*{Date}&Corner&\multicolumn{2}{c}{2}& \multicolumn{2}{c}{4}&\multicolumn{2}{c}{13}& \multicolumn{2}{c}{14} \\
 \cmidrule(r){3-4}  \cmidrule(r){5-6} \cmidrule(r){7-8} \cmidrule(r){9-10}
& Method& mean& $\sigma$ & mean& $\sigma$&  mean& $\sigma$ & mean& $\sigma$\\ [0.5ex]
\hline\noalign{\smallskip} 
\multirow{2}*{2016}&GD1&(-365.724,4010.315)&(0.022,0.058)&(-362.840,-20.635)&(0.020,0.040)&(15.129,-144.431)&(0.011,0.032)&(4110.078,-132.512)&(0.018,0.037)\\
&GD2&(-365.835,4010.479)&(0.023,0.039)&(-362.864,-20.706)&(0.021,0.031)&(15.108,-144.558)&(0.020,0.035)&(4110.114,-132.581)&(0.023,0.029)\\
\multirow{2}*{2017}&GD1&(-364.492,4010.520)&(0.036,0.032)&(-362.200,-20.438)&(0.031,0.025)&(13.533,-141.278)&(0.019,0.031)&(4108.507,-133.841)&(0.028,0.050)\\
&GD2&(-364.520,4010.478)&(0.026,0.029)&(-362.197,-20.513)&(0.050,0.043)&(13.560,-141.323)&(0.034,0.047)&(4108.603,-133.934)&(0.035,0.068)\\
\hline 
\end{tabular}
\tablefoot{GD1 means the results of photographic astrometry and GD2 means the results of differential astrometry. Column 3 to 10 show the mean pixel positions and their corresponding precisions. The top table shows the results taking CCD$\#$2 as the reference and the bottom table shows the results taking CCD$\#$3 as the reference. The averages of the differences between the results of photographic astrometry and differential astrometry are about 0.018 pixels~($\sim$0.008 arcsec) for Obs16 and about 0.026 pixels~($\sim$0.012 arcsec) for Obs17 respectively. And the average precisions are about 0.049 pixels~($\sim$0.022 arcsec) for Obs16 and about 0.056 pixels~($\sim$0.025 arcsec) for Obs17 respectively. All units are in pixels.}
\label{tab4}
\end{sidewaystable}

For comparison with the AK03 method in the next section, we consider the average of the transformed positions as the average gap between chips. The relative angle between chips can be estimated by the slope of a fitted straight line of the transformed positions. The results are shown in Table~\ref{tab5}.

For the two types of astrometry, there are some little differences between them in rotation angle, which are within 0.002 degrees. For Obs16, the average gaps derived from photographic astrometry and differential astrometry differ by about 0.046 pixel~($\sim$0.021 arcsec). For Obs17, the average gaps derived from photographic astrometry and differential astrometry only differ by about 0.001 pixel. The precisions of the gaps derived from the two types of astrometry achieve a comparable level. The average precisions of the gaps are about 0.018 pixel~($\sim$0.008 arcsec) for Obs16 and 0.028 pixel~($\sim$0.013 arcsec) for Obs17, respectively.

\begin{table*}[htb]
\centering
\caption[]{Statistics of the average gaps and the relative angles between chips.}
\small
 \begin{tabular}{clccccccccc}
 \\
  \hline\noalign{\smallskip}
 Date&Item&  $<$mean$>$& $\sigma$ & $<$mean$>$& $\sigma$&  $<$mean$>$& $\sigma$ & $<$mean$>$& $\sigma$\\
 &&\multicolumn{2}{c}{CCD1(CCD2)}& \multicolumn{2}{c}{CCD4(CCD3)}&\multicolumn{2}{c}{CCD4(CCD2)}& \multicolumn{2}{c}{CCD1(CCD3)} \\
  \hline\noalign{\smallskip}
\multirow{4}*{2016}&GD1gap&142.853&0.019&138.590&0.018&373.893&0.023&364.373&0.014\\
&GD1angle&-0.003&0.001&0.167&0.001&89.877&0.000&89.960&0.000\\
&GD2gap&142.935&0.017&138.649&0.022&373.925&0.016&364.385&0.015\\
&GD2angle&-0.002&0.001&0.169&0.001&89.877&0.001&89.960&0.000\\
  \hline\noalign{\smallskip}
 \multirow{4}*{2017}&GD1gap&142.079&0.029&137.686&0.028&369.456&0.021&363.432&0.023\\
&GD1angle&0.023&0.001&0.104&0.001&89.904&0.001&89.968&0.001\\
&GD2gap&142.091&0.034&137.682&0.035&369.460&0.028&363.417&0.028\\
&GD2angle&0.022&0.001&0.104&0.001&89.903&0.001&89.968&0.001\\  \hline
\end{tabular}
\tablefoot{GD1 means the result of photographic astrometry and GD2 means the result of differential astrometry. Columns 3 to 10 show the average and the precision for the gaps and the angles. The reference chips are in parentheses. The results of the gaps are in pixels and the results of relative angles are in degrees.}
\label{tab5}
\end{table*}

\section{Comparison with the AK03 method}
As our method is implemented by transforming the adjacent edge of one chip to the actual pixel coordinate of the reference chip, it is a physical metric for the CCD mosaic chips. In contrast, the AK03 method determines the geometry of CCD mosaic chips when each chip is put into a distortion-free coordinate. Although the AK03 method mainly aims to offer convenient solutions for many applications needing to transform the GD-corrected positions of each chip into a common distortion-free reference frame, it is still useful for monitoring the progress of the interchip offset of CCD mosaic chips. In this section, we make a comparison between the two methods from this perspective.

Of the CCD\#k, the distortion-corrected positions ($x_{k}^{\rm corr},y_{k}^{\rm corr}$) were transformed to the distortion-corrected positions of reference chip~(CCD\#ref,ref$\neq$k) using a six-parameter linear transform, which can absorb the DAR in the following first-order terms:
$$
\begin{array}{rcl}
\left(
\begin{array}{c}
x_{ref}^{\rm corr}\\
y_{ref}^{\rm corr}\\
\end{array}
\right)
&\!=\!& \!
\left[
\begin{array}{cc}
\!\! A & B \\
\!\! C & D \\
\end{array}
\!\right]
\!\left(
\begin{array}{c}
x_{k}^{\rm corr}\\
y_{k}^{\rm corr}\\
\end{array}
\!\right)
+
\left(
\begin{array}{c}
\!x_{k0}\\
\!y_{k0}\\
\end{array}
\!\right).
\end{array}
$$

The relative scale between CCD\#k and CCD\#ref can be derived by $\alpha\textnormal{=}\sqrt{AD\textnormal{-}BC}$ and the rotation angle $\theta\textnormal{=}\arctan(B\textnormal{-}C,A\textnormal{+}D)$ according to \citet{Anderson2007}. For each frame, we can proceed to solve the relative quantities~(scale, relative angle, and gap) when CCD\#2 and CCD\#3 are taken as references, respectively. The results are shown in Appendix B, from Table~\ref{tab7} to Table~\ref{tab10}.

\begin{figure*}[htp]
\centering
\includegraphics[width=0.48\textwidth]{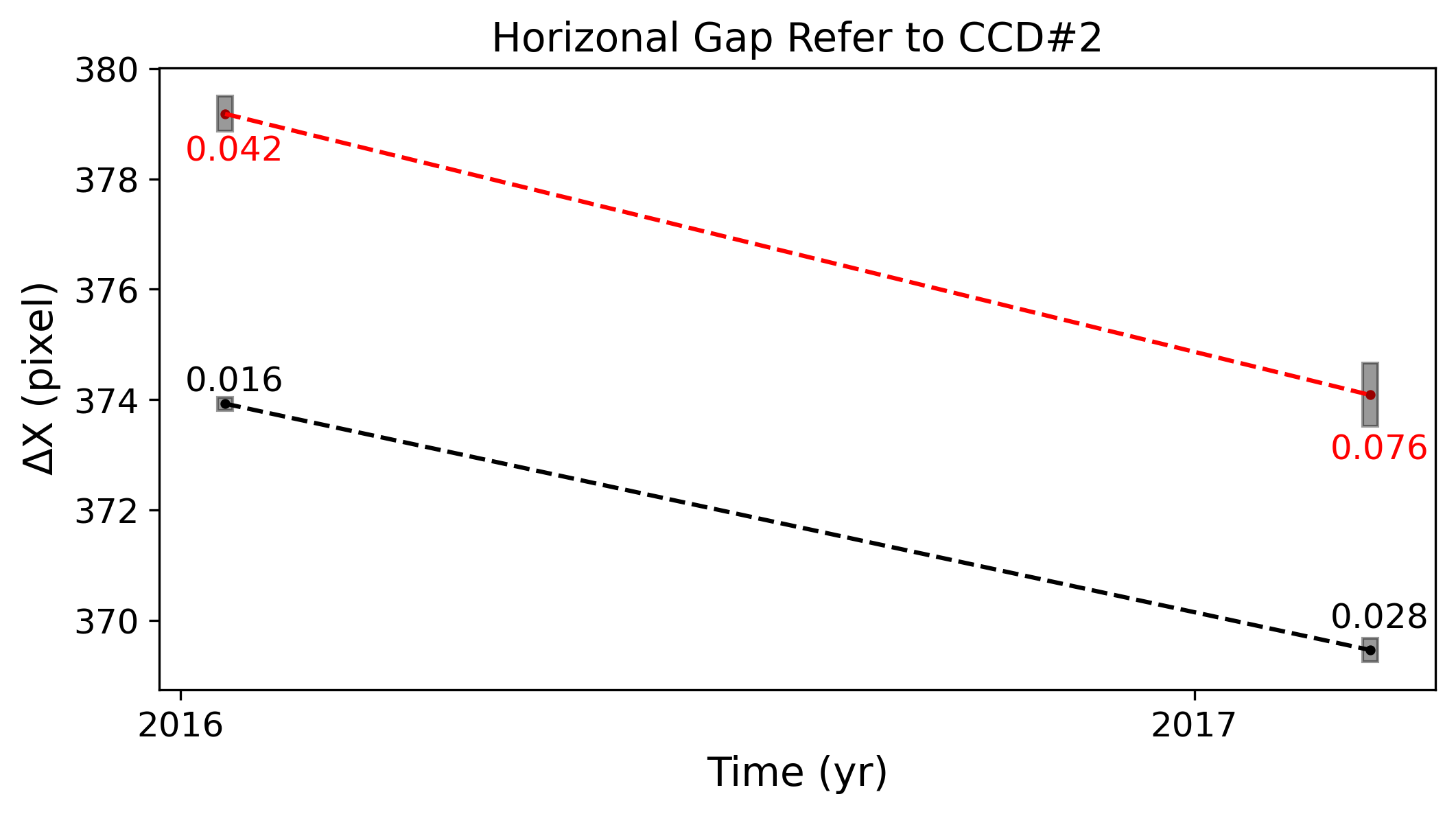}
\includegraphics[width=0.48\textwidth]{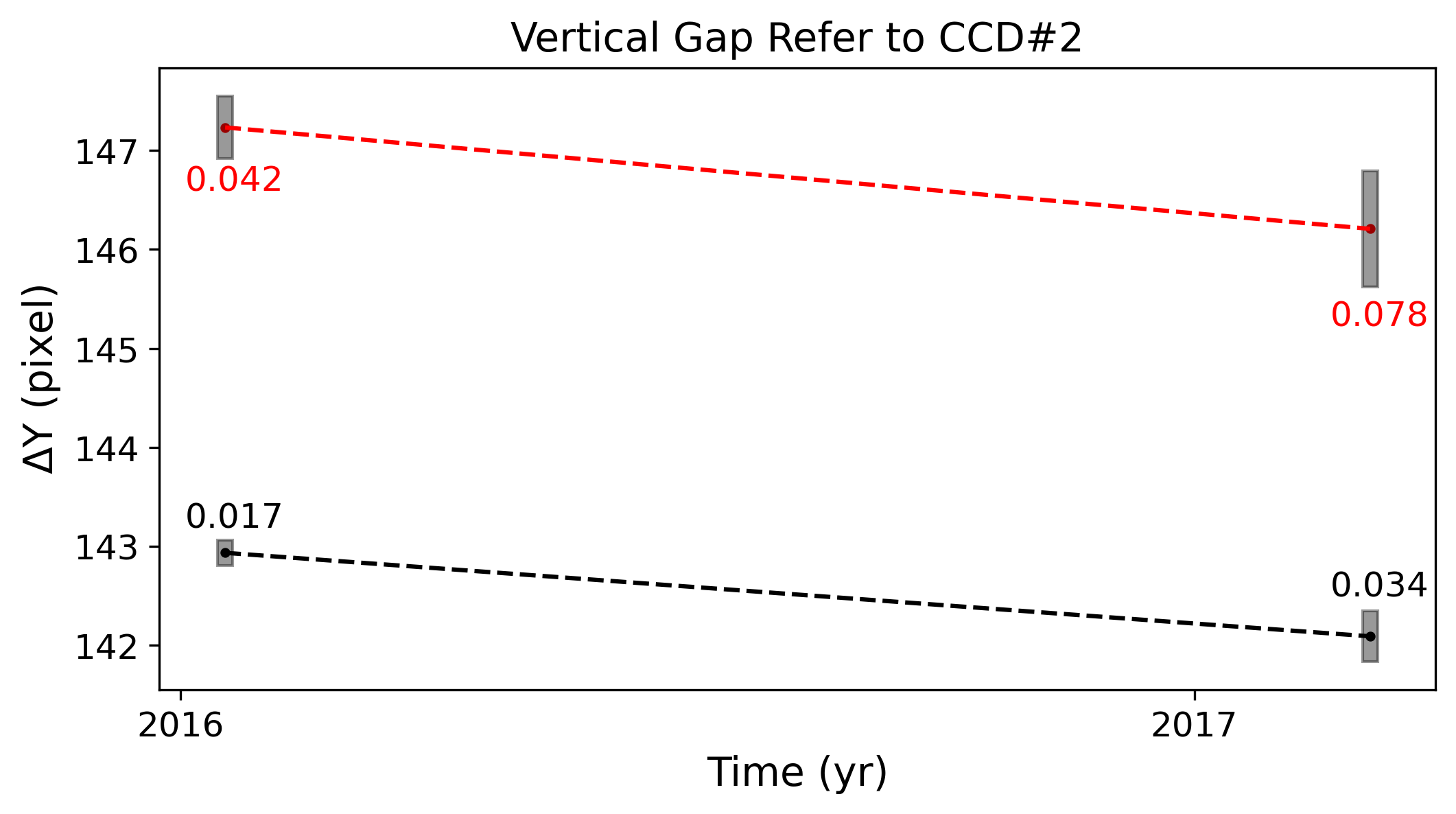}
\caption{Averages and precisions~($\sigma$) of the gap in the horizonal and vertical directions as a function of time, when CCD\#2 is taken as reference. We note that the gaps of 2016 are based on measurements with the DES r filter and the gaps of 2017 are based on measurements with the SDSS g filter. The black points represent the results derived from the proposed method in this paper, and the red points represent the results derived from the AK03 method. The shaded bars indicate a $\pm\sigma\times15$ (for clarity) interval around the mean gap location. The values of $\sigma$ are shown in the panel.}
\label{ccd14refccd2}
\end{figure*}

\begin{figure*}[htp]
\centering
\includegraphics[width=0.48\textwidth]{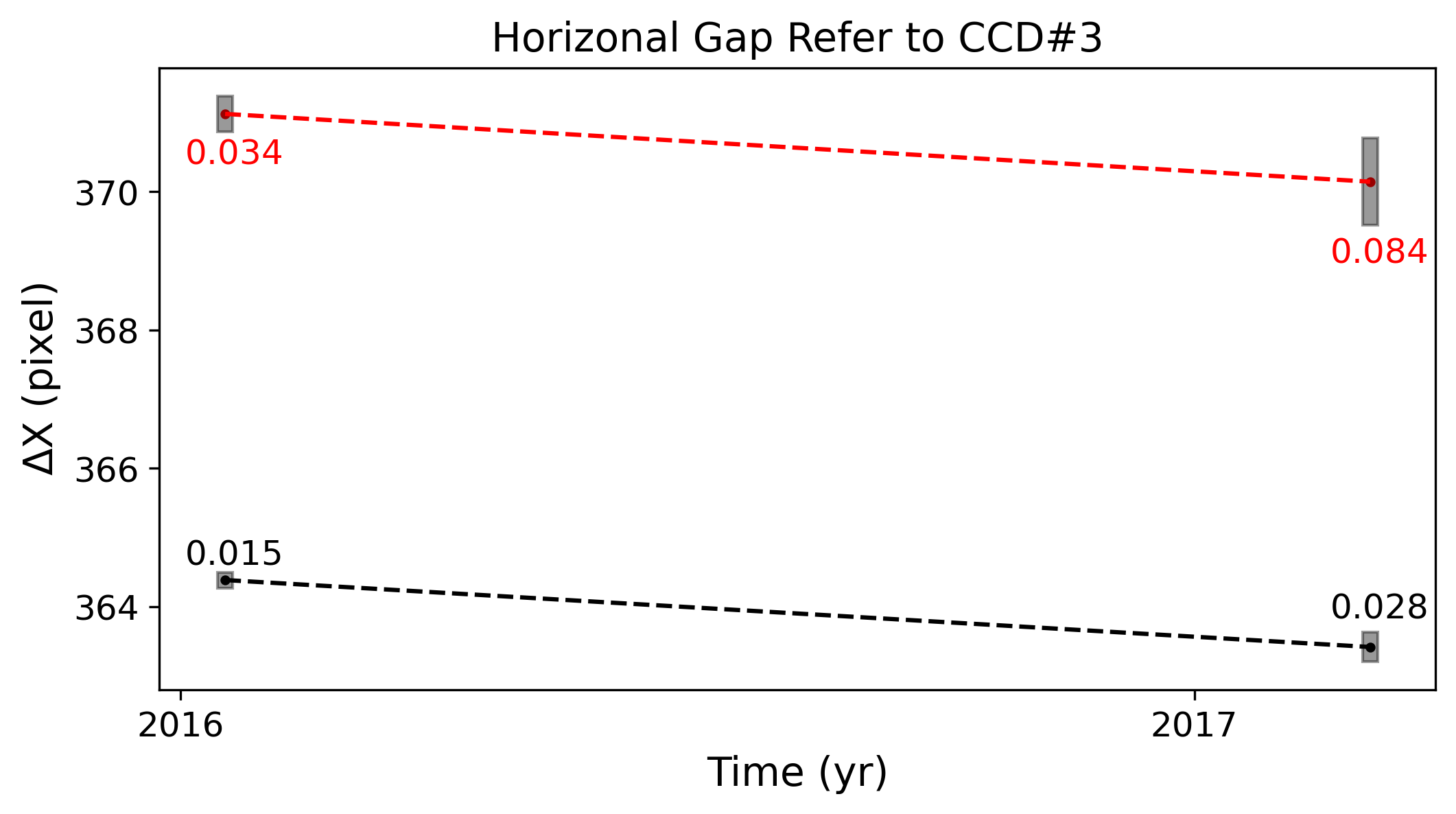}
\includegraphics[width=0.48\textwidth]{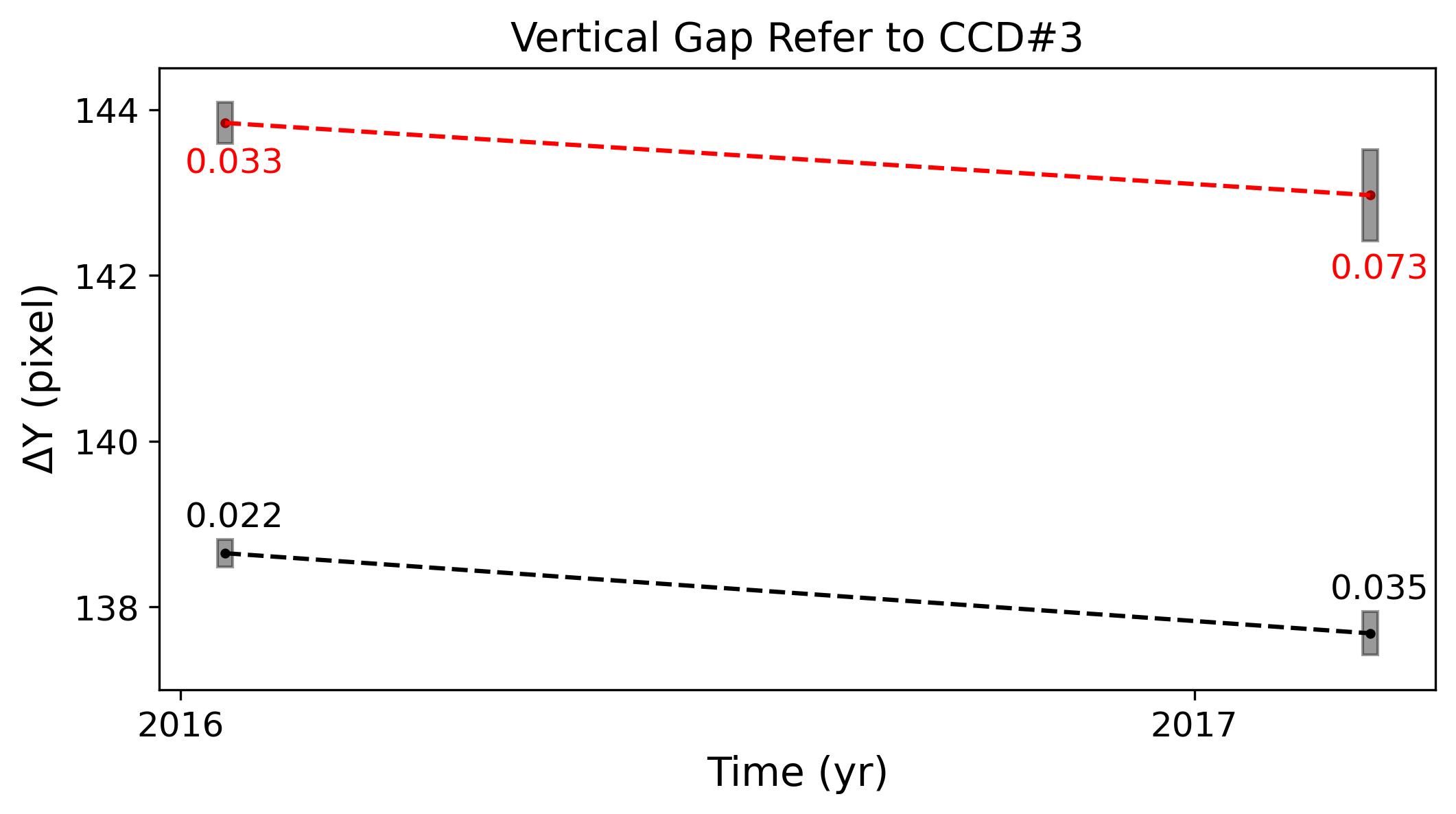}
\caption{Similar to Fig.~\ref{ccd14refccd2}. The variations of the gap in the horizonal and vertical direction as a function of time are shown, with CCD\#3 taken as reference. We note that the gaps of 2016 are based on measurements with the DES r filter and the gaps of 2017 are based on measurements with the SDSS g filter.}
\label{ccd14refccd3}
\end{figure*}

Because our method is implemented in actual or physical coordinate including GD effects, while the AK03 method is implemented in a distortion-free coordinate, we only compare the precisions of the relative angle and gap of the two methods. For the relative angle between chips, the two methods have similar precisions. For the relative gap, we plot their values and precisions of the two epochs's observations in Fig.~\ref{ccd14refccd2} and Fig.~\ref{ccd14refccd3}, although they are based on measurements with different filters. We note that the results of our method are derived from differential astrometry. Similar changes in the offset can be detected by both methods; however, the solution proposed in this paper shows at least a factor of two improvement in precision on average over the AK03 method. We think there are two definite advantages to our method. On one hand, we perform the measurements for two adjacent edges instead of two individual chips, making the results as local as possible and meanwhile alleviating the propagated error of the transformation and the GD model throughout the FOV. On the other hand, the final outcome is not mixed with the GD effects, which would bias the realistic geometry of the CCD mosaic chips.

\section{Conclusion}
In this paper, we propose a solution to determine the actual geometry of CCD mosaic chips by taking advantage of the GD solution derived from the observations. Taking into consideration the fact that there may be few stars available for astrometric calibration during the deep observation of large ground-based or space-base telescope, we referred to the idea from the astrometry of the HST to only use stars' pixel positions to derive the relative positions between chips. We refer to the practice of only using the pixel position as differential astrometry in this paper. In order to ensure the results are reliable, we took advantage of Gaia EDR3 to derive the relative positions between chips, as was done in \cite{Wang2019}, to provide a close comparison. We refer to the practice of using an astrometric catalogue as photographic astrometry.

We implemented the technique for the CCD mosaic chips of the Bok 2.3-m telescope at Kitt Peak, based on two epochs of observations~(Jan 17,2016 and Mar 5,2017). For the two epochs of observations, the average gaps derived from photographic astrometry and differential astrometry differ by about 0.046 pixels~($\sim$0.021 arcsec) and 0.001 pixels~(<0.001 arcsec), respectively, while the average precisions of the gaps are about 0.018 pixels~($\sim$0.008 arcsec) and 0.028 pixels~($\sim$0.013 arcsec), respectively.

We also made a comparison with the AK03 method, which is adopted for measuring the geometry of WFPC2 chips and WFC3 chips at the HST. The results show that the solution proposed in this paper shows an improvement of at least a factor of two in precision on average.

This paper shows there is a good agreement between the two types of astrometry for the relative positions between chips. This is important for the deep observations of the planned CSST. Since the CSST can observe much fainter stars~(e.g. 26 mag in the g band) than the Gaia satellite, of which we only have the pixel positions rather than their astrometric parameters~(e.g. J2000.0 positions, proper motions, parallaxes, etc.), differential astrometry is expected to be more suitable for determining the relative positions between chips. In the mean time, the internal positional precisions can be substantially improved by precisely transforming the positions from different chips into a global coordinate system. However, accurate solutions for the GD and the geometry of CCD mosaic chips are needed. So, we recommend verifying the calibration on a regular basis, since the experiences with the HST show the importance of detecting long-term variations of the GD~\citep{2007Variation,2014ACS}, which are caused by, for example, the slow out-gassing of metals~\citep{Bedin2018MNRAS}. This would also lead to long-term variations of the geometry of CCD mosaic chips.
\begin{acknowledgements}
This work was supported by the China Manned Space Project with NO.~CMS-CSST-2021-B08, by the National Natural Science Foundation of China (Grant Nos.~11873026,~11273014), by
the Joint Research Fund in Astronomy (Grant No. U1431227) under
cooperative agreement between the National Natural Science
Foundation of China~(NSFC) and Chinese Academy Sciences~(CAS), and
partly by the Fundamental Research Funds for the Central
Universities and Excellent Postgraduate Recommendation Scientific Research Innovative Cultivation Program of Jinan University. This work has made use of data from the European Space Agency (ESA) mission \emph{Gaia} (\url{https://www.cosmos.esa.int/gaia}), processed by the \emph{Gaia} Data Processing and Analysis Consortium (DPAC, \url{https://www.cosmos.esa.int/web/gaia/dpac/consortium}). Funding for the DPAC has been provided by national institutions, in particular the institutions participating in the \emph{Gaia} Multilateral Agreement.

\end{acknowledgements}

\bibliographystyle{aa}
\bibliography{bibtex}

\begin{thebibliography}{28}
\expandafter\ifx\csname natexlab\endcsname\relax\def\natexlab#1{#1}\fi

\bibitem[{{Anderson}(2007)}]{Anderson2007}
{Anderson}, J. 2007, {Variation of the Distortion Solution}, Instrument Science
  Report ACS 2007-08

\bibitem[{Anderson(2007)}]{2007Variation}
Anderson, J. 2007, Proceedings of the American Mathematical Society

\bibitem[{{Anderson} \& {King}(2003)}]{Anderson2003}
{Anderson}, J. \& {King}, I.~R. 2003, \pasp, 115, 113

\bibitem[{{Bedin} \& {Fontanive}(2018)}]{Bedin2018MNRAS}
{Bedin}, L.~R. \& {Fontanive}, C. 2018, \mnras, 481, 5339

\bibitem[{{Bellini} \& {Bedin}(2009)}]{Bellini2009}
{Bellini}, A. \& {Bedin}, L.~R. 2009, \pasp, 121, 1419

\bibitem[{{Bernstein} {et~al.}(2017){Bernstein}, {Armstrong}, {Plazas},
  {Walker}, {Abbott}, {Allam}, {Bechtol}, {Benoit-L{\'e}vy}, {Brooks}, {Burke},
  {Carnero Rosell}, {Carrasco Kind}, {Carretero}, {Cunha}, {da Costa}, {DePoy},
  {Desai}, {Diehl}, {Eifler}, {Fernandez}, {Fosalba}, {Frieman},
  {Garc{\'\i}a-Bellido}, {Gerdes}, {Gruen}, {Gruendl}, {Gschwend}, {Gutierrez},
  {Honscheid}, {James}, {Kent}, {Krause}, {Kuehn}, {Kuropatkin}, {Li}, {Maia},
  {March}, {Marshall}, {Menanteau}, {Miquel}, {Ogando}, {Reil}, {Roodman},
  {Rykoff}, {Sanchez}, {Scarpine}, {Schindler}, {Schubnell}, {Sevilla-Noarbe},
  {Smith}, {Smith}, {Soares-Santos}, {Sobreira}, {Suchyta}, {Swanson}, {Tarle},
  \& {DES Collaboration}}]{Bernstein2017}
{Bernstein}, G.~M., {Armstrong}, R., {Plazas}, A.~A., {et~al.} 2017, \pasp,
  129, 074503

\bibitem[{{DePoy} {et~al.}(2008){DePoy}, {Abbott}, {Annis}, {Antonik},
  {Barcel{\'o}}, {Bernstein}, {Bigelow}, {Brooks}, {Buckley-Geer}, {Campa},
  {Cardiel}, {Castander}, {Castilla}, {Cease}, {Chappa}, {Dede}, {Derylo},
  {Diehl}, {Doel}, {DeVicente}, {Estrada}, {Finley}, {Flaugher}, {Gaztanaga},
  {Gerdes}, {Gladders}, {Guarino}, {Gutierrez}, {Hamilton}, {Haney}, {Holland},
  {Honscheid}, {Huffman}, {Karliner}, {Kau}, {Kent}, {Kozlovsky}, {Kubik},
  {Kuehn}, {Kuhlmann}, {Kuk}, {Leger}, {Lin}, {Martinez}, {Martinez},
  {Merritt}, {Mohr}, {Moore}, {Moore}, {Nord}, {Ogando}, {Olsen}, {Onal},
  {Peoples}, {Qian}, {Roe}, {Sanchez}, {Scarpine}, {Schmidt}, {Schmitt},
  {Schubnell}, {Schultz}, {Selen}, {Shaw}, {Simaitis}, {Slaughter}, {Smith},
  {Spinka}, {Stefanik}, {Stuermer}, {Talaga}, {Tarle}, {Thaler}, {Tucker},
  {Walker}, {Worswick}, \& {Zhao}}]{DePoy2008}
{DePoy}, D.~L., {Abbott}, T., {Annis}, J., {et~al.} 2008, in Society of
  Photo-Optical Instrumentation Engineers (SPIE) Conference Series, Vol. 7014,
  Ground-based and Airborne Instrumentation for Astronomy II, 70140E

\bibitem[{{Fabricius} {et~al.}(2021){Fabricius}, {Luri}, {Arenou}, {Babusiaux},
  {Helmi}, {Muraveva}, {Reyl{\'e}}, {Spoto}, {Vallenari}, {Antoja}, {Balbinot},
  {Barache}, {Bauchet}, {Bragaglia}, {Busonero}, {Cantat-Gaudin}, {Carrasco},
  {Diakit{\'e}}, {Fabrizio}, {Figueras}, {Garcia-Gutierrez}, {Garofalo},
  {Jordi}, {Kervella}, {Khanna}, {Leclerc}, {Licata}, {Lambert}, {Marrese},
  {Masip}, {Ramos}, {Robichon}, {Robin}, {Romero-G{\'o}mez}, {Rubele}, \&
  {Weiler}}]{Fabricius2021}
{Fabricius}, C., {Luri}, X., {Arenou}, F., {et~al.} 2021, \aap, 649, A5

\bibitem[{{French} {et~al.}(2006){French}, {McGhee}, {Frey}, {Hock}, {Rounds},
  {Jacobson}, \& {Verbiscer}}]{French2006}
{French}, R.~G., {McGhee}, C.~A., {Frey}, M., {et~al.} 2006, \pasp, 118, 246

\bibitem[{{Gaia Collaboration} {et~al.}(2018){Gaia Collaboration}, {Brown},
  {Vallenari}, {Prusti}, {de Bruijne}, {Babusiaux}, {Bailer-Jones}, {Biermann},
  {Evans}, {Eyer}, {Jansen}, {Jordi}, {Klioner}, {Lammers}, {Lindegren},
  {Luri}, {Mignard}, {Panem}, {Pourbaix}, {Randich}, {Sartoretti}, {Siddiqui},
  {Soubiran}, {van Leeuwen}, {Walton}, {Arenou}, {Bastian}, {Cropper},
  {Drimmel}, {Katz}, {Lattanzi}, {Bakker}, {Cacciari}, {Casta{\~n}eda},
  {Chaoul}, {Cheek}, {De Angeli}, {Fabricius}, {Guerra}, {Holl}, {Masana},
  {Messineo}, {Mowlavi}, {Nienartowicz}, {Panuzzo}, {Portell}, {Riello},
  {Seabroke}, {Tanga}, {Th{\'e}venin}, {Gracia-Abril}, {Comoretto},
  {Garcia-Reinaldos}, {Teyssier}, {Altmann}, {Andrae}, {Audard},
  {Bellas-Velidis}, {Benson}, {Berthier}, {Blomme}, {Burgess}, {Busso},
  {Carry}, {Cellino}, {Clementini}, {Clotet}, {Creevey}, {Davidson}, {De
  Ridder}, {Delchambre}, {Dell'Oro}, {Ducourant},
  {Fern{\'a}ndez-Hern{\'a}ndez}, {Fouesneau}, {Fr{\'e}mat}, {Galluccio},
  {Garc{\'\i}a-Torres}, {Gonz{\'a}lez-N{\'u}{\~n}ez}, {Gonz{\'a}lez-Vidal},
  {Gosset}, {Guy}, {Halbwachs}, {Hambly}, {Harrison}, {Hern{\'a}ndez},
  {Hestroffer}, {Hodgkin}, {Hutton}, {Jasniewicz}, {Jean-Antoine-Piccolo},
  {Jordan}, {Korn}, {Krone-Martins}, {Lanzafame}, {Lebzelter}, {L{\"o}ffler},
  {Manteiga}, {Marrese}, {Mart{\'\i}n-Fleitas}, {Moitinho}, {Mora}, {Muinonen},
  {Osinde}, {Pancino}, {Pauwels}, {Petit}, {Recio-Blanco}, {Richards},
  {Rimoldini}, {Robin}, {Sarro}, {Siopis}, {Smith}, {Sozzetti}, {S{\"u}veges},
  {Torra}, {van Reeven}, {Abbas}, {Abreu Aramburu}, {Accart}, {Aerts},
  {Altavilla}, {{\'A}lvarez}, {Alvarez}, {Alves}, {Anderson}, {Andrei},
  {Anglada Varela}, {Antiche}, {Antoja}, {Arcay}, {Astraatmadja}, {Bach},
  {Baker}, {Balaguer-N{\'u}{\~n}ez}, {Balm}, {Barache}, {Barata}, {Barbato},
  {Barblan}, {Barklem}, {Barrado}, {Barros}, {Barstow}, {Bartholom{\'e}
  Mu{\~n}oz}, {Bassilana}, {Becciani}, {Bellazzini}, {Berihuete}, {Bertone},
  {Bianchi}, {Bienaym{\'e}}, {Blanco-Cuaresma}, {Boch}, {Boeche}, {Bombrun},
  {Borrachero}, {Bossini}, {Bouquillon}, {Bourda}, {Bragaglia}, {Bramante},
  {Breddels}, {Bressan}, {Brouillet}, {Br{\"u}semeister}, {Brugaletta},
  {Bucciarelli}, {Burlacu}, {Busonero}, {Butkevich}, {Buzzi}, {Caffau},
  {Cancelliere}, {Cannizzaro}, {Cantat-Gaudin}, {Carballo}, {Carlucci},
  {Carrasco}, {Casamiquela}, {Castellani}, {Castro-Ginard}, {Charlot},
  {Chemin}, {Chiavassa}, {Cocozza}, {Costigan}, {Cowell}, {Crifo}, {Crosta},
  {Crowley}, {Cuypers}, {Dafonte}, {Damerdji}, {Dapergolas}, {David}, {David},
  {de Laverny}, {De Luise}, {De March}, {de Martino}, {de Souza}, {de Torres},
  {Debosscher}, {del Pozo}, {Delbo}, {Delgado}, {Delgado}, {Di Matteo},
  {Diakite}, {Diener}, {Distefano}, {Dolding}, {Drazinos}, {Dur{\'a}n},
  {Edvardsson}, {Enke}, {Eriksson}, {Esquej}, {Eynard Bontemps}, {Fabre},
  {Fabrizio}, {Faigler}, {Falc{\~a}o}, {Farr{\`a}s Casas}, {Federici},
  {Fedorets}, {Fernique}, {Figueras}, {Filippi}, {Findeisen}, {Fonti},
  {Fraile}, {Fraser}, {Fr{\'e}zouls}, {Gai}, {Galleti}, {Garabato},
  {Garc{\'\i}a-Sedano}, {Garofalo}, {Garralda}, {Gavel}, {Gavras}, {Gerssen},
  {Geyer}, {Giacobbe}, {Gilmore}, {Girona}, {Giuffrida}, {Glass}, {Gomes},
  {Granvik}, {Gueguen}, {Guerrier}, {Guiraud}, {Guti{\'e}rrez-S{\'a}nchez},
  {Haigron}, {Hatzidimitriou}, {Hauser}, {Haywood}, {Heiter}, {Helmi}, {Heu},
  {Hilger}, {Hobbs}, {Hofmann}, {Holland}, {Huckle}, {Hypki}, {Icardi},
  {Jan{\ss}en}, {Jevardat de Fombelle}, {Jonker}, {Juh{\'a}sz}, {Julbe},
  {Karampelas}, {Kewley}, {Klar}, {Kochoska}, {Kohley}, {Kolenberg},
  {Kontizas}, {Kontizas}, {Koposov}, {Kordopatis}, {Kostrzewa-Rutkowska},
  {Koubsky}, {Lambert}, {Lanza}, {Lasne}, {Lavigne}, {Le Fustec}, {Le
  Poncin-Lafitte}, {Lebreton}, {Leccia}, {Leclerc}, {Lecoeur-Taibi},
  {Lenhardt}, {Leroux}, {Liao}, {Licata}, {Lindstr{\o}m}, {Lister}, {Livanou},
  {Lobel}, {L{\'o}pez}, {Managau}, {Mann}, {Mantelet}, {Marchal}, {Marchant},
  {Marconi}, {Marinoni}, {Marschalk{\'o}}, {Marshall}, {Martino}, {Marton},
  {Mary}, {Massari}, {Matijevi{\v{c}}}, {Mazeh}, {McMillan}, {Messina},
  {Michalik}, {Millar}, {Molina}, {Molinaro}, {Moln{\'a}r}, {Montegriffo},
  {Mor}, {Morbidelli}, {Morel}, {Morris}, {Mulone}, {Muraveva}, {Musella},
  {Nelemans}, {Nicastro}, {Noval}, {O'Mullane}, {Ord{\'e}novic},
  {Ord{\'o}{\~n}ez-Blanco}, {Osborne}, {Pagani}, {Pagano}, {Pailler},
  {Palacin}, {Palaversa}, {Panahi}, {Pawlak}, {Piersimoni}, {Pineau}, {Plachy},
  {Plum}, {Poggio}, {Poujoulet}, {Pr{\v{s}}a}, {Pulone}, {Racero}, {Ragaini},
  {Rambaux}, {Ramos-Lerate}, {Regibo}, {Reyl{\'e}}, {Riclet}, {Ripepi}, {Riva},
  {Rivard}, {Rixon}, {Roegiers}, {Roelens}, {Romero-G{\'o}mez}, {Rowell},
  {Royer}, {Ruiz-Dern}, {Sadowski}, {Sagrist{\`a} Sell{\'e}s}, {Sahlmann},
  {Salgado}, {Salguero}, {Sanna}, {Santana-Ros}, {Sarasso}, {Savietto},
  {Schultheis}, {Sciacca}, {Segol}, {Segovia}, {S{\'e}gransan}, {Shih},
  {Siltala}, {Silva}, {Smart}, {Smith}, {Solano}, {Solitro}, {Sordo}, {Soria
  Nieto}, {Souchay}, {Spagna}, {Spoto}, {Stampa}, {Steele},
  {Steidelm{\"u}ller}, {Stephenson}, {Stoev}, {Suess}, {Surdej}, {Szabados},
  {Szegedi-Elek}, {Tapiador}, {Taris}, {Tauran}, {Taylor}, {Teixeira},
  {Terrett}, {Teyssandier}, {Thuillot}, {Titarenko}, {Torra Clotet}, {Turon},
  {Ulla}, {Utrilla}, {Uzzi}, {Vaillant}, {Valentini}, {Valette}, {van Elteren},
  {Van Hemelryck}, {van Leeuwen}, {Vaschetto}, {Vecchiato}, {Veljanoski},
  {Viala}, {Vicente}, {Vogt}, {von Essen}, {Voss}, {Votruba}, {Voutsinas},
  {Walmsley}, {Weiler}, {Wertz}, {Wevers}, {Wyrzykowski}, {Yoldas},
  {{\v{Z}}erjal}, {Ziaeepour}, {Zorec}, {Zschocke}, {Zucker}, {Zurbach}, \&
  {Zwitter}}]{Gaia2018}
{Gaia Collaboration}, {Brown}, A.~G.~A., {Vallenari}, A., {et~al.} 2018, \aap,
  616, A1

\bibitem[{{Gaia Collaboration} {et~al.}(2021){Gaia Collaboration}, {Brown},
  {Vallenari}, {Prusti}, {de Bruijne}, {Babusiaux}, {Biermann}, {Creevey},
  {Evans}, {Eyer}, {Hutton}, {Jansen}, {Jordi}, {Klioner}, {Lammers},
  {Lindegren}, {Luri}, {Mignard}, {Panem}, {Pourbaix}, {Randich}, {Sartoretti},
  {Soubiran}, {Walton}, {Arenou}, {Bailer-Jones}, {Bastian}, {Cropper},
  {Drimmel}, {Katz}, {Lattanzi}, {van Leeuwen}, {Bakker}, {Cacciari},
  {Casta{\~n}eda}, {De Angeli}, {Ducourant}, {Fabricius}, {Fouesneau},
  {Fr{\'e}mat}, {Guerra}, {Guerrier}, {Guiraud}, {Jean-Antoine Piccolo},
  {Masana}, {Messineo}, {Mowlavi}, {Nicolas}, {Nienartowicz}, {Pailler},
  {Panuzzo}, {Riclet}, {Roux}, {Seabroke}, {Sordo}, {Tanga}, {Th{\'e}venin},
  {Gracia-Abril}, {Portell}, {Teyssier}, {Altmann}, {Andrae}, {Bellas-Velidis},
  {Benson}, {Berthier}, {Blomme}, {Brugaletta}, {Burgess}, {Busso}, {Carry},
  {Cellino}, {Cheek}, {Clementini}, {Damerdji}, {Davidson}, {Delchambre},
  {Dell'Oro}, {Fern{\'a}ndez-Hern{\'a}ndez}, {Galluccio}, {Garc{\'\i}a-Lario},
  {Garcia-Reinaldos}, {Gonz{\'a}lez-N{\'u}{\~n}ez}, {Gosset}, {Haigron},
  {Halbwachs}, {Hambly}, {Harrison}, {Hatzidimitriou}, {Heiter},
  {Hern{\'a}ndez}, {Hestroffer}, {Hodgkin}, {Holl}, {Jan{\ss}en}, {Jevardat de
  Fombelle}, {Jordan}, {Krone-Martins}, {Lanzafame}, {L{\"o}ffler}, {Lorca},
  {Manteiga}, {Marchal}, {Marrese}, {Moitinho}, {Mora}, {Muinonen}, {Osborne},
  {Pancino}, {Pauwels}, {Petit}, {Recio-Blanco}, {Richards}, {Riello},
  {Rimoldini}, {Robin}, {Roegiers}, {Rybizki}, {Sarro}, {Siopis}, {Smith},
  {Sozzetti}, {Ulla}, {Utrilla}, {van Leeuwen}, {van Reeven}, {Abbas}, {Abreu
  Aramburu}, {Accart}, {Aerts}, {Aguado}, {Ajaj}, {Altavilla}, {{\'A}lvarez},
  {{\'A}lvarez Cid-Fuentes}, {Alves}, {Anderson}, {Anglada Varela}, {Antoja},
  {Audard}, {Baines}, {Baker}, {Balaguer-N{\'u}{\~n}ez}, {Balbinot}, {Balog},
  {Barache}, {Barbato}, {Barros}, {Barstow}, {Bartolom{\'e}}, {Bassilana},
  {Bauchet}, {Baudesson-Stella}, {Becciani}, {Bellazzini}, {Bernet}, {Bertone},
  {Bianchi}, {Blanco-Cuaresma}, {Boch}, {Bombrun}, {Bossini}, {Bouquillon},
  {Bragaglia}, {Bramante}, {Breedt}, {Bressan}, {Brouillet}, {Bucciarelli},
  {Burlacu}, {Busonero}, {Butkevich}, {Buzzi}, {Caffau}, {Cancelliere},
  {C{\'a}novas}, {Cantat-Gaudin}, {Carballo}, {Carlucci}, {Carnerero},
  {Carrasco}, {Casamiquela}, {Castellani}, {Castro-Ginard}, {Castro Sampol},
  {Chaoul}, {Charlot}, {Chemin}, {Chiavassa}, {Cioni}, {Comoretto}, {Cooper},
  {Cornez}, {Cowell}, {Crifo}, {Crosta}, {Crowley}, {Dafonte}, {Dapergolas},
  {David}, {David}, {de Laverny}, {De Luise}, {De March}, {De Ridder}, {de
  Souza}, {de Teodoro}, {de Torres}, {del Peloso}, {del Pozo}, {Delbo},
  {Delgado}, {Delgado}, {Delisle}, {Di Matteo}, {Diakite}, {Diener},
  {Distefano}, {Dolding}, {Eappachen}, {Edvardsson}, {Enke}, {Esquej}, {Fabre},
  {Fabrizio}, {Faigler}, {Fedorets}, {Fernique}, {Fienga}, {Figueras},
  {Fouron}, {Fragkoudi}, {Fraile}, {Franke}, {Gai}, {Garabato},
  {Garcia-Gutierrez}, {Garc{\'\i}a-Torres}, {Garofalo}, {Gavras}, {Gerlach},
  {Geyer}, {Giacobbe}, {Gilmore}, {Girona}, {Giuffrida}, {Gomel}, {Gomez},
  {Gonzalez-Santamaria}, {Gonz{\'a}lez-Vidal}, {Granvik},
  {Guti{\'e}rrez-S{\'a}nchez}, {Guy}, {Hauser}, {Haywood}, {Helmi}, {Hidalgo},
  {Hilger}, {H{\l}adczuk}, {Hobbs}, {Holland}, {Huckle}, {Jasniewicz},
  {Jonker}, {Juaristi Campillo}, {Julbe}, {Karbevska}, {Kervella}, {Khanna},
  {Kochoska}, {Kontizas}, {Kordopatis}, {Korn}, {Kostrzewa-Rutkowska},
  {Kruszy{\'n}ska}, {Lambert}, {Lanza}, {Lasne}, {Le Campion}, {Le Fustec},
  {Lebreton}, {Lebzelter}, {Leccia}, {Leclerc}, {Lecoeur-Taibi}, {Liao},
  {Licata}, {Lindstr{\o}m}, {Lister}, {Livanou}, {Lobel}, {Madrero Pardo},
  {Managau}, {Mann}, {Marchant}, {Marconi}, {Marcos Santos}, {Marinoni},
  {Marocco}, {Marshall}, {Martin Polo}, {Mart{\'\i}n-Fleitas}, {Masip},
  {Massari}, {Mastrobuono-Battisti}, {Mazeh}, {McMillan}, {Messina},
  {Michalik}, {Millar}, {Mints}, {Molina}, {Molinaro}, {Moln{\'a}r},
  {Montegriffo}, {Mor}, {Morbidelli}, {Morel}, {Morris}, {Mulone}, {Munoz},
  {Muraveva}, {Murphy}, {Musella}, {Noval}, {Ord{\'e}novic}, {Orr{\`u}},
  {Osinde}, {Pagani}, {Pagano}, {Palaversa}, {Palicio}, {Panahi}, {Pawlak},
  {Pe{\~n}alosa Esteller}, {Penttil{\"a}}, {Piersimoni}, {Pineau}, {Plachy},
  {Plum}, {Poggio}, {Poretti}, {Poujoulet}, {Pr{\v{s}}a}, {Pulone}, {Racero},
  {Ragaini}, {Rainer}, {Raiteri}, {Rambaux}, {Ramos}, {Ramos-Lerate}, {Re
  Fiorentin}, {Regibo}, {Reyl{\'e}}, {Ripepi}, {Riva}, {Rixon}, {Robichon},
  {Robin}, {Roelens}, {Rohrbasser}, {Romero-G{\'o}mez}, {Rowell}, {Royer},
  {Rybicki}, {Sadowski}, {Sagrist{\`a} Sell{\'e}s}, {Sahlmann}, {Salgado},
  {Salguero}, {Samaras}, {Sanchez Gimenez}, {Sanna}, {Santove{\~n}a},
  {Sarasso}, {Schultheis}, {Sciacca}, {Segol}, {Segovia}, {S{\'e}gransan},
  {Semeux}, {Shahaf}, {Siddiqui}, {Siebert}, {Siltala}, {Slezak}, {Smart},
  {Solano}, {Solitro}, {Souami}, {Souchay}, {Spagna}, {Spoto}, {Steele},
  {Steidelm{\"u}ller}, {Stephenson}, {S{\"u}veges}, {Szabados}, {Szegedi-Elek},
  {Taris}, {Tauran}, {Taylor}, {Teixeira}, {Thuillot}, {Tonello}, {Torra},
  {Torra}, {Turon}, {Unger}, {Vaillant}, {van Dillen}, {Vanel}, {Vecchiato},
  {Viala}, {Vicente}, {Voutsinas}, {Weiler}, {Wevers}, {Wyrzykowski}, {Yoldas},
  {Yvard}, {Zhao}, {Zorec}, {Zucker}, {Zurbach}, \& {Zwitter}}]{Gaia2021}
{Gaia Collaboration}, {Brown}, A.~G.~A., {Vallenari}, A., {et~al.} 2021, \aap,
  649, A1

\bibitem[{{Gubler} \& {Tytler}(1998)}]{Gubler1998}
{Gubler}, J. \& {Tytler}, D. 1998, \pasp, 110, 738

\bibitem[{{Gunn} {et~al.}(1998){Gunn}, {Carr}, {Rockosi}, {Sekiguchi}, {Berry},
  {Elms}, {de Haas}, {Ivezi{\'c}}, {Knapp}, {Lupton}, {Pauls}, {Simcoe},
  {Hirsch}, {Sanford}, {Wang}, {York}, {Harris}, {Annis}, {Bartozek},
  {Boroski}, {Bakken}, {Haldeman}, {Kent}, {Holm}, {Holmgren}, {Petravick},
  {Prosapio}, {Rechenmacher}, {Doi}, {Fukugita}, {Shimasaku}, {Okada}, {Hull},
  {Siegmund}, {Mannery}, {Blouke}, {Heidtman}, {Schneider}, {Lucinio}, \&
  {Brinkman}}]{Gunn1998}
{Gunn}, J.~E., {Carr}, M., {Rockosi}, C., {et~al.} 1998, \aj, 116, 3040

\bibitem[{{Lin} {et~al.}(2020){Lin}, {Peng}, \& {Zheng}}]{Lin2020}
{Lin}, F.~R., {Peng}, Q.~Y., \& {Zheng}, Z.~J. 2020, \mnras, 498, 258

\bibitem[{{Lou} {et~al.}(2016){Lou}, {Liang}, {Yao}, {Zheng}, {Cheng}, {Wang},
  {Liu}, {Qian}, {Zhao}, \& {Yang}}]{Lou2016}
{Lou}, Z., {Liang}, M., {Yao}, D., {et~al.} 2016, in Society of Photo-Optical
  Instrumentation Engineers (SPIE) Conference Series, Vol. 10154, Society of
  Photo-Optical Instrumentation Engineers (SPIE) Conference Series, 101542A

\bibitem[{{Luppino} {et~al.}(1995){Luppino}, {Mezger}, \&
  {Miyazaki}}]{Luppino1995}
{Luppino}, G.~A., {Mezger}, M.~R., \& {Miyazaki}, S. 1995, in IAU Symposium,
  Vol. 167, New Developments in Array Technology and Applications, ed. A.~G.~D.
  {Philip}, K.~{Janes}, \& A.~R. {Upgren}, 297

\bibitem[{{Peng} {et~al.}(2017){Peng}, {Peng}, \& {Wang}}]{Peng2017}
{Peng}, H.~W., {Peng}, Q.~Y., \& {Wang}, N. 2017, \mnras, 467, 2266

\bibitem[{{Peng} {et~al.}(2012){Peng}, {Vienne}, {Zhang}, {Desmars}, {Yang}, \&
  {He}}]{Peng2012}
{Peng}, Q.~Y., {Vienne}, A., {Zhang}, Q.~F., {et~al.} 2012, \aj, 144, 170

\bibitem[{{Peng} {et~al.}(2015){Peng}, {Wang}, {Vienne}, {Zhang}, {Li}, \&
  {Meng}}]{Peng2015MNRAS}
{Peng}, Q.~Y., {Wang}, N., {Vienne}, A., {et~al.} 2015, \mnras, 449, 2638

\bibitem[{{Platais} {et~al.}(2002){Platais}, {Kozhurina-Platais}, {Girard},
  {van Altena}, {Klemola}, {Stauffer}, {Armandroff}, {Mighell}, {Dell'Antonio},
  {Falco}, \& {Sarajedini}}]{Platais2002}
{Platais}, I., {Kozhurina-Platais}, V., {Girard}, T.~M., {et~al.} 2002, \aj,
  124, 601

\bibitem[{{Riello} {et~al.}(2021){Riello}, {De Angeli}, {Evans}, {Montegriffo},
  {Carrasco}, {Busso}, {Palaversa}, {Burgess}, {Diener}, {Davidson}, {Rowell},
  {Fabricius}, {Jordi}, {Bellazzini}, {Pancino}, {Harrison}, {Cacciari}, {van
  Leeuwen}, {Hambly}, {Hodgkin}, {Osborne}, {Altavilla}, {Barstow}, {Brown},
  {Castellani}, {Cowell}, {De Luise}, {Gilmore}, {Giuffrida}, {Hidalgo},
  {Holland}, {Marinoni}, {Pagani}, {Piersimoni}, {Pulone}, {Ragaini}, {Rainer},
  {Richards}, {Sanna}, {Walton}, {Weiler}, \& {Yoldas}}]{Riello2021}
{Riello}, M., {De Angeli}, F., {Evans}, D.~W., {et~al.} 2021, \aap, 649, A3

\bibitem[{{Sekiguchi} {et~al.}(1992){Sekiguchi}, {Iwashita}, {Doi},
  {Kashikawa}, \& {Okamura}}]{Sekiguchi1992}
{Sekiguchi}, M., {Iwashita}, H., {Doi}, M., {Kashikawa}, N., \& {Okamura}, S.
  1992, \pasp, 104, 744

\bibitem[{Ubeda {et~al.}(2014)Ubeda, Kozhurina-Platais, \& Bedin}]{2014ACS}
Ubeda, L., Kozhurina-Platais, V., \& Bedin, L.~R. 2014, Instrument Science
  Report ACS 2013-03, 12 pages

\bibitem[{{Wadadekar} {et~al.}(2006){Wadadekar}, {Casertano}, {Hook},
  {K{\i}z{\i}ltan}, {Koekemoer}, {Ferguson}, \& {Denchev}}]{Wadadekar2006}
{Wadadekar}, Y., {Casertano}, S., {Hook}, R., {et~al.} 2006, \pasp, 118, 450

\bibitem[{{Wang} {et~al.}(2017){Wang}, {Peng}, {Peng}, {Xie}, {Ma}, \&
  {Zhang}}]{Wang2017}
{Wang}, N., {Peng}, Q.~Y., {Peng}, H.~W., {et~al.} 2017, \mnras, 468, 1415

\bibitem[{{Wang} {et~al.}(2019){Wang}, {Peng}, {Zhou}, {Peng}, \&
  {Peng}}]{Wang2019}
{Wang}, N., {Peng}, Q.~Y., {Zhou}, X., {Peng}, X.~Y., \& {Peng}, H.~W. 2019,
  \mnras, 485, 1626

\bibitem[{{Zhan}(2021)}]{ZhanHu2021}
{Zhan}, H. 2021, Chinese Science Bulletin, 66, 1290

\bibitem[{Zheng {et~al.}(2021)Zheng, Peng, \& Lin}]{Zheng2021}
Zheng, Z.~J., Peng, Q.~Y., \& Lin, F.~R. 2021, \mnras, 502, 6216

\end{thebibliography}

\begin{appendix}
\section{Third-order polynomial for the GD solutions}

\begin{table*}[htp]
\caption{Coefficients and their errors of the third-order polynomial for each chip, derived from Obs16.}
\label{tab_GD16}
\centering
\begin{tabular}{cccccccccc}
\noalign{\smallskip}
\hline
\hline
\noalign{\smallskip}
Term $\!(k)\!\!\!\!\!$&Polyn.$\!\!\!\!\!\!\!\!\!$
&$a_{k}$&$b_{k}$&$\sigma_{a}$&$\sigma_{b}$  &$a_{k}$&$b_{k}$&$\sigma_{a}$&$\sigma_{b}$ \\
\noalign{\smallskip}\hline
\noalign{\smallskip}
& & \multicolumn{8}{c}  {CCD\#1} \\
\noalign{\smallskip}
& & \multicolumn{4}{c}  {GD1} & \multicolumn{4}{c}  {GD2}\\
\noalign{\smallskip}
0&$\tilde{x}^{0}(\tilde{y}^{0})$&1.782&-1.876&0.002&0.005&1.947&-2.194&0.003&0.008
\\
1&$\tilde{x}$            &0.559&-4.617&0.005&0.014&-0.235&-4.347&0.008&0.019
\\
2&$\tilde{y}$ &-4.048&-0.291&0.005&0.014&-3.852&-1.099&0.008&0.019
\\
3&$\tilde{x}^2$&-6.298&1.980&0.003&0.010&-6.145&2.076&0.005&0.014
\\
4&$\tilde{x}\tilde{y}$&3.897&-4.199&0.003&0.008&3.775&-3.882&0.005&0.012
\\
5&$\tilde{y}^2$&-2.047&5.666&0.003&0.010&-2.221&5.635&0.005&0.014
\\
6&$\tilde{x}^3$&1.917&-0.020&0.006&0.019&1.926&-0.025&0.011&0.027
\\
7&$\tilde{x}^2\tilde{y}$&-0.098&1.888&0.006&0.017&-0.096&1.883&0.009&0.024
\\
8&$\tilde{x}\tilde{y}^2$&1.871&-0.088&0.006&0.017&1.887&-0.100&0.009&0.024
\\
9&$\tilde{y}^3$&-0.026&2.140&0.006&0.019&-0.027&2.269&0.011&0.027
\\
\noalign{\smallskip}
\hline
\noalign{\smallskip}
& & \multicolumn{8}{c}  {CCD\#2} \\
\noalign{\smallskip}
& & \multicolumn{4}{c}  {GD1} & \multicolumn{4}{c}  {GD2}\\
\noalign{\smallskip}
0&$\tilde{x}^{0}(\tilde{y}^{0})$&1.729&1.479&0.001&0.003&1.862&2.045&0.002&0.004
\\
1&$\tilde{x}$            &0.836&4.055&0.003&0.007&-0.100&4.406&0.005&0.010
\\
2&$\tilde{y}$ &4.100&-0.288&0.003&0.007&4.303&-1.209&0.005&0.010
\\
3&$\tilde{x}^2$&-6.303&-1.838&0.002&0.005&-6.154&-1.741&0.004&0.007
\\
4&$\tilde{x}\tilde{y}$&-3.622&-4.191&0.002&0.004&-3.821&-3.881&0.003&0.006
\\
5&$\tilde{y}^2$&-2.038&-5.305&0.002&0.005&-2.210&-5.423&0.004&0.007
\\
6&$\tilde{x}^3$&1.911&0.029&0.005&0.01&1.912&0.022&0.008&0.014
\\
7&$\tilde{x}^2\tilde{y}$&0.069&1.871&0.004&0.009&0.072&1.870&0.007&0.012
\\
8&$\tilde{x}\tilde{y}^2$&1.857&0.067&0.004&0.009&1.875&0.074&0.007&0.012
\\
9&$\tilde{y}^3$&0.024&1.892&0.005&0.010&0.008&1.968&0.008&0.014
\\
\noalign{\smallskip}
\hline
\noalign{\smallskip}
& & \multicolumn{8}{c}  {CCD\#3} \\
\noalign{\smallskip}
& & \multicolumn{4}{c}  {GD1} & \multicolumn{4}{c}  {GD2}\\
\noalign{\smallskip}
0&$\tilde{x}^{0}(\tilde{y}^{0})$&-1.578&-1.566&0.002&0.003&-1.968&-2.111&0.002&0.004
\\
1&$\tilde{x}$            &0.298&4.006&0.004&0.008&-0.745&4.609&0.006&0.011
\\
2&$\tilde{y}$ &4.066&0.349&0.004&0.008&4.073&-0.657&0.006&0.011
\\
3&$\tilde{x}^2$&5.841&1.994&0.003&0.006&5.995&2.083&0.004&0.008
\\
4&$\tilde{x}\tilde{y}$&3.898&3.857&0.003&0.005&3.785&4.166&0.004&0.007
\\
5&$\tilde{y}^2$&1.898&5.721&0.003&0.006&1.769&5.684&0.004&0.008
\\
6&$\tilde{x}^3$&1.894&0.036&0.006&0.011&1.898&0.032&0.009&0.015
\\
7&$\tilde{x}^2\tilde{y}$&0.096&1.894&0.005&0.010&0.099&1.904&0.008&0.013
\\
8&$\tilde{x}\tilde{y}^2$&1.858&0.099&0.005&0.010&1.878&0.091&0.008&0.013
\\
9&$\tilde{y}^3$&0.041&1.900&0.006&0.011&0.040&1.979&0.009&0.015
\\
\noalign{\smallskip}
\hline
\noalign{\smallskip}
& & \multicolumn{8}{c}  {CCD\#4} \\
\noalign{\smallskip}
& & \multicolumn{4}{c}  {GD1} & \multicolumn{4}{c}  {GD2}\\
\noalign{\smallskip}
0&$\tilde{x}^{0}(\tilde{y}^{0})$&-1.423&1.352&0.002&0.003&-1.880&2.106&0.003&0.005
\\
1&$\tilde{x}$            &0.838&-3.805&0.004&0.008&-0.302&-3.751&0.007&0.012
\\
2&$\tilde{y}$ &-3.637&0.168&0.004&0.008&-3.170&-0.923&0.007&0.012
\\
3&$\tilde{x}^2$&5.875&-1.847&0.003&0.006&6.024&-1.756&0.005&0.008
\\
4&$\tilde{x}\tilde{y}$&-3.622&3.894&0.003&0.005&-3.815&4.205&0.004&0.007
\\
5&$\tilde{y}^2$&1.926&-5.285&0.003&0.006&1.799&-5.409&0.005&0.008
\\
6&$\tilde{x}^3$&1.902&-0.037&0.006&0.012&1.900&-0.047&0.009&0.017
\\
7&$\tilde{x}^2\tilde{y}$&-0.098&1.892&0.005&0.010&-0.099&1.903&0.008&0.014
\\
8&$\tilde{x}\tilde{y}^2$&1.871&-0.101&0.005&0.010&1.891&-0.111&0.008&0.014
\\
9&$\tilde{y}^3$&-0.032&1.933&0.006&0.012&-0.035&2.019&0.009&0.017
\\
\noalign{\smallskip}
\hline
\end{tabular}
\end{table*}

\begin{table*}[htp]
\caption{Coefficients and their errors of the third-order polynomial for each chip, derived from Obs17.}
\label{tab_GD17}
\centering
\begin{tabular}{cccccccccc}
\noalign{\smallskip}
\hline
\hline
\noalign{\smallskip}
Term $\!(k)\!\!\!\!\!$&Polyn.$\!\!\!\!\!\!\!\!\!$
&$a_{k}$&$b_{k}$&$\sigma_{a}$&$\sigma_{b}$  &$a_{k}$&$b_{k}$&$\sigma_{a}$&$\sigma_{b}$ \\
\noalign{\smallskip}\hline
\noalign{\smallskip}
& & \multicolumn{8}{c}  {CCD\#1} \\
\noalign{\smallskip}
& & \multicolumn{4}{c}  {GD1} & \multicolumn{4}{c}  {GD2}\\
\noalign{\smallskip}
0&$\tilde{x}^{0}(\tilde{y}^{0})$&1.706&-1.594&0.001&0.001&2.677&-2.354&0.001&0.001
\\
1&$\tilde{x}$            &0.699&-4.336&0.002&0.003&-0.853&-4.303&0.003&0.003
\\
2&$\tilde{y}$ &-4.459&0.235&0.002&0.003&-3.830&-0.659&0.003&0.003
\\
3&$\tilde{x}^2$&-6.310&2.008&0.001&0.002&-6.399&2.158&0.002&0.002
\\
4&$\tilde{x}\tilde{y}$&3.966&-4.194&0.001&0.002&3.744&-4.297&0.002&0.002
\\
5&$\tilde{y}^2$&-2.041&5.800&0.001&0.002&-2.022&5.709&0.002&0.002
\\
6&$\tilde{x}^3$&1.934&-0.021&0.003&0.004&1.952&-0.010&0.004&0.005
\\
7&$\tilde{x}^2\tilde{y}$&-0.085&1.887&0.002&0.003&-0.090&1.895&0.003&0.004
\\
8&$\tilde{x}\tilde{y}^2$&1.854&-0.074&0.002&0.003&1.873&-0.077&0.003&0.004
\\
9&$\tilde{y}^3$&-0.012&1.889&0.003&0.004&-0.015&1.919&0.004&0.005
\\
\noalign{\smallskip}
\hline
\noalign{\smallskip}
& & \multicolumn{8}{c}  {CCD\#2} \\
\noalign{\smallskip}
& & \multicolumn{4}{c}  {GD1} & \multicolumn{4}{c}  {GD2}\\
\noalign{\smallskip}
0&$\tilde{x}^{0}(\tilde{y}^{0})$&1.487&1.358&0.001&0.001&3.249&2.081&0.001&0.001
\\
1&$\tilde{x}$            &1.283&3.983&0.002&0.002&-0.631&4.357&0.003&0.003
\\
2&$\tilde{y}$ &4.022&0.044&0.002&0.002&4.424&-1.207&0.003&0.003
\\
3&$\tilde{x}^2$&-6.294&-1.813&0.002&0.002&-6.377&-1.666&0.002&0.002
\\
4&$\tilde{x}\tilde{y}$&-3.581&-4.149&0.001&0.001&-3.901&-4.257&0.002&0.002
\\
5&$\tilde{y}^2$&-2.025&-5.276&0.002&0.002&-2.001&-5.461&0.002&0.002
\\
6&$\tilde{x}^3$&1.920&0.009&0.003&0.003&1.937&0.017&0.004&0.004
\\
7&$\tilde{x}^2\tilde{y}$&0.030&1.879&0.003&0.003&0.032&1.891&0.004&0.004
\\
8&$\tilde{x}\tilde{y}^2$&1.865&0.028&0.003&0.003&1.881&0.032&0.004&0.004
\\
9&$\tilde{y}^3$&0.002&1.869&0.003&0.003&0.003&1.895&0.004&0.004
\\
\noalign{\smallskip}
\hline
\noalign{\smallskip}
& & \multicolumn{8}{c}  {CCD\#3} \\
\noalign{\smallskip}
& & \multicolumn{4}{c}  {GD1} & \multicolumn{4}{c}  {GD2}\\
\noalign{\smallskip}
0&$\tilde{x}^{0}(\tilde{y}^{0})$&-1.578&-1.566&0.002&0.003&-1.968&-2.111&0.002&0.004
\\
1&$\tilde{x}$            &0.298&4.006&0.004&0.008&-0.745&4.609&0.006&0.011
\\
2&$\tilde{y}$ &4.066&0.349&0.004&0.008&4.073&-0.657&0.006&0.011
\\
3&$\tilde{x}^2$&5.841&1.994&0.003&0.006&5.995&2.083&0.004&0.008
\\
4&$\tilde{x}\tilde{y}$&3.898&3.857&0.003&0.005&3.785&4.166&0.004&0.007
\\
5&$\tilde{y}^2$&1.898&5.721&0.003&0.006&1.769&5.684&0.004&0.008
\\
6&$\tilde{x}^3$&1.894&0.036&0.006&0.011&1.898&0.032&0.009&0.015
\\
7&$\tilde{x}^2\tilde{y}$&0.096&1.894&0.005&0.010&0.099&1.904&0.008&0.013
\\
8&$\tilde{x}\tilde{y}^2$&1.858&0.099&0.005&0.010&1.878&0.091&0.008&0.013
\\
9&$\tilde{y}^3$&0.041&1.900&0.006&0.011&0.040&1.979&0.009&0.015
\\
\noalign{\smallskip}
\hline
\noalign{\smallskip}
& & \multicolumn{8}{c}  {CCD\#4} \\
\noalign{\smallskip}
& & \multicolumn{4}{c}  {GD1} & \multicolumn{4}{c}  {GD2}\\
\noalign{\smallskip}
0&$\tilde{x}^{0}(\tilde{y}^{0})$&-1.423&1.352&0.002&0.003&-1.880&2.106&0.003&0.005
\\
1&$\tilde{x}$            &0.300&4.171&0.002&0.002&-1.583&4.809&0.003&0.003
\\
2&$\tilde{y}$ &4.226&0.504&0.002&0.002&4.411&-0.739&0.003&0.003
\\
3&$\tilde{x}^2$&5.971&2.041&0.001&0.002&5.994&2.180&0.002&0.002
\\
4&$\tilde{x}\tilde{y}$&3.982&3.912&0.001&0.001&3.745&3.891&0.002&0.002
\\
5&$\tilde{y}^2$&1.925&5.846&0.001&0.002&1.952&5.742&0.002&0.002
\\
6&$\tilde{x}^3$&1.918&0.009&0.003&0.003&1.930&-0.004&0.004&0.004
\\
7&$\tilde{x}^2\tilde{y}$&0.037&1.859&0.002&0.003&0.033&1.873&0.003&0.004
\\
8&$\tilde{x}\tilde{y}^2$&1.828&0.040&0.002&0.003&1.845&0.037&0.003&0.004
\\
9&$\tilde{y}^3$&0.017&1.820&0.003&0.003&0.017&1.838&0.004&0.004
\\
\noalign{\smallskip}
\hline
\end{tabular}
\end{table*}
\section{Average relative quantities between CCD chips derived from the AK03 method}
\begin{table}[htp]
\centering
\caption{Average relative quantities~(scale, relative angle and gap) derived from AK03 method for Obs16 when CCD\#2 is taken as a reference, with formal errors.}
\begin{tabular}{cccc}
 & & &  \\
\hline
\hline
 & & &  \\
Parameter &       $k=$[1] &  $k=$[3] &  $k=$[4]\\
 & & &  \\
\hline
 & & &  \\
$\alpha_{k}/\alpha_{[2]}$&0.99997&0.99987&1.00004\\
                &$\pm$0.00001& $\pm$0.00002& $\pm$0.00001\\
 & & &  \\
$\theta_{k}$$-$$\theta_{[2]}$&0.029&0.067&$-$0.106\\
                &\textbf{$\pm$0.001}&$\pm$0.001&\textbf{$\pm$0.001}\\
& & &  \\
$(x_{[2]}^{\rm corr})_k$&$-$7.701&4459.920&4475.180\\
                &$\pm$0.050&$\pm$0.063 &\textbf{$\pm$0.042}\\
& & &  \\
$(y_{[2]}^{\rm corr})_k$&4179.232& 4196.955& 21.632\\
                &\textbf{$\pm$0.042}&$\pm$0.096&$\pm$0.028\\
 & & &  \\
\hline
\end{tabular}
\tablefoot{The values for $\theta_{k}$$-$$\theta_{[2]}$ are expressed in degrees. The relative shift $(x_{[2]}^{\rm corr})_k$ and $(y_{[2]}^{\rm corr})_k$ are given in pixels with regard to the system of CCD\#2. The formal errors in bold are compared with the ones derived from our method.}
\label{tab7}
\end{table}

\begin{table}[htp]
\centering
\caption{Similar quantities~(scale, relative angle and gap) for Obs16 to those in Table~\ref{tab7} when CCD\#3 is taken as a reference, with formal errors.}
\begin{tabular}{cccc}
 & & &  \\
\hline
\hline
 & & &  \\
Parameter &       $k=$[1] & $k=$[2] &  $k=$[4]\\
 & & &  \\
\hline
 & & &  \\
$\alpha_{k}/\alpha_{[3]}$&1.00010&1.00013&1.00017\\
                &$\pm$0.00002& $\pm$0.00002&$\pm$0.00001\\
 & & &  \\
$\theta_{k}$$-$$\theta_{[3]}$&$-$0.038&$-$0.067&$-$0.171\\
                &\textbf{$\pm$0.000}&$\pm$0.001&\textbf{$\pm$0.001}\\
& & &  \\
$(x_{[3]}^{\rm corr})_k$&$-$4467.127&$-$4454.508&21.187\\
                &\textbf{$\pm$0.034}& $\pm$0.059&$\pm$0.050\\
& & &  \\
$(y_{[3]}^{\rm corr})_k$&$-$22.911&$-$4202.644&$-$4175.838\\
                &$\pm$0.035&$\pm$0.079&\textbf{$\pm$0.033}\\
 & & &  \\
\hline
\end{tabular}
\tablefoot{The values for $\theta_{k}$$-$$\theta_{[3]}$ are expressed in degrees. The relative shift $(x_{[3]}^{\rm corr})_k$ and $(y_{[3]}^{\rm corr})_k$ are given in pixels with regard to the system of CCD\#3. The formal errors in bold are compared with the ones derived from our method.}
\label{tab8}
\end{table}

\begin{table}[htp]
\centering
\caption{Average relative quantities~(scale, relative angle and gap) derived from AK03 method for Obs17 when CCD\#2 is taken as a reference, with formal errors.}
\begin{tabular}{cccc}
 & & &  \\
\hline
\hline
 & & &  \\
Parameter &       $k=$[1] &  $k=$[3] &  $k=$[4]\\
 & & &  \\
\hline
 & & &  \\
$\alpha_{k}/\alpha_{[2]}$&0.99997&1.00006&1.00021\\
                &$\pm$0.00002& $\pm$0.00003& $\pm$0.00003\\
 & & &  \\
$\theta_{k}$$-$$\theta_{[2]}$&$-$0.024&0.015&$-$0.074\\
                &\textbf{$\pm$0.001}&$\pm$0.001&\textbf{$\pm$0.002}\\
& & &  \\
$(x_{[2]}^{\rm corr})_k$&$-$11.772&4454.720&4470.083\\
                &$\pm$0.046&$\pm$0.109 &\textbf{$\pm$0.076}\\
& & &  \\
$(y_{[2]}^{\rm corr})_k$&4178.207& 4199.352& 23.171\\
                &\textbf{$\pm$0.078}&$\pm$0.146&$\pm$0.092\\
 & & &  \\
\hline
\end{tabular}
\tablefoot{The values for $\theta_{k}$$-$$\theta_{[2]}$ are expressed in degrees. The relative shift $(x_{[2]}^{\rm corr})_k$ and $(y_{[2]}^{\rm corr})_k$ are given in pixels with regard to the system of CCD\#2. The formal errors in bold are compared with the ones derived from our method.}
\label{tab9}
\end{table}

\begin{table}[htp]
\centering
\caption{Similar quantities~(scale, relative angle and gap) for Obs17 to those in Table~\ref{tab9} when CCD\#3 is taken as a reference, with formal errors.}
\begin{tabular}{cccc}
 & & &  \\
\hline
\hline
 & & &  \\
Parameter &       $k=$[1] & $k=$[2] &  $k=$[4]\\
 & & &  \\
\hline
 & & &  \\
$\alpha_{k}/\alpha_{[3]}$&0.99990&0.99994&1.00015\\
                &$\pm$0.00003& $\pm$0.00003&$\pm$0.00002\\
 & & &  \\
$\theta_{k}$$-$$\theta_{[3]}$&$-$0.039&$-$0.015&$-$0.088\\
                &\textbf{$\pm$0.001}&$\pm$0.001&\textbf{$\pm$0.001}\\
& & &  \\
$(x_{[3]}^{\rm corr})_k$&$-$4466.147&$-$4452.300&16.439\\
                &\textbf{$\pm$0.084}& $\pm$0.114&$\pm$0.056\\
& & &  \\
$(y_{[3]}^{\rm corr})_k$&$-$22.285&$-$4200.284&$-$4174.967\\
                &$\pm$0.070&$\pm$0.093&\textbf{$\pm$0.073}\\
 & & &  \\
\hline
\end{tabular}
\tablefoot{The values for $\theta_{k}$$-$$\theta_{[3]}$ are expressed in degrees. The relative shift $(x_{[3]}^{\rm corr})_k$ and $(y_{[3]}^{\rm corr})_k$ are given in pixels with regard to the system of CCD\#3. The formal errors in bold are compared with the ones derived from our method.}
\label{tab10}
\end{table}
\end{appendix}
\end{document}